\documentclass[acmsmall]{acmart}

\AtBeginDocument{%
  \providecommand\BibTeX{{%
    \normalfont B\kern-0.5em{\scshape i\kern-0.25em b}\kern-0.8em\TeX}}}

\setcopyright{acmlicensed}
\acmJournal{TORS}
\acmYear{2024} 
\acmVolume{xxx} \acmNumber{xxx} \acmArticle{} 
\acmMonth{10} 
\acmDOI{10.1145/3701763}

\newcommand{\ie}{\emph{i.e., }}
\newcommand{\eg}{\emph{e.g., }}

\newcommand{\wrt}{\emph{w.r.t. }}
\newcommand{\cf}{\emph{cf. }}

\usepackage{enumitem}
\usepackage{balance}
\usepackage{verbatim}
\usepackage[ruled]{algorithm2e}
\usepackage{multirow}
\usepackage{subfigure} 
\usepackage{ragged2e}
\usepackage{caption}
\usepackage{graphicx}
\usepackage[normalem]{ulem}
\useunder{\uline}{\ul}{}
\usepackage{amsmath}
\usepackage{setspace}
\usepackage{bm}
\usepackage{longtable}
\begin{document}

\title{Recommendation Unlearning via Influence Function}
% \title{Unveiling Enhanced Recommendation Unlearning through Influence Functions}

\author{Yang Zhang}
\orcid{0000-0002-7863-5183}
\affiliation{
  \institution{University of Science and Technology of China}
  \city{Hefei}
  \country{China}
}
\email{zyang1580@gmail.com}
\authornote{
Currently affiliated with the National University of Singapore.
}

\author{Zhiyu Hu}
\orcid{0009-0004-9092-8888}
\affiliation{
  \institution{University of Science and Technology of China}
  \city{Hefei}
  \country{China}}
\email{zhiyuhu@mail.ustc.edu.cn}

\author{Yimeng Bai}
\orcid{0009-0008-8874-9409}
\affiliation{
  \institution{University of Science and Technology of China}
  \city{Hefei}
  \country{China}}
\email{baiyimeng@mail.ustc.edu.cn}

\author{Jiancan Wu}
\orcid{0000-0002-6941-5218}
\affiliation{
  \department{Electronic Engineering and Information Science}
  \institution{University of Science and Technology of China}
  \city{Hefei}
  \country{China}}
\email{wujcan@gmail.com}

\author{Qifan Wang}
\orcid{0000-0002-7570-5756}
\affiliation{
  \department{Meta AI}
  \institution{Meta Platforms Inc}
  \city{Menlo Park}
  \country{United States}}
\email{wqfcr@fb.com}

\author{Fuli Feng}
\orcid{0000-0002-5828-9842}
\affiliation{
  \department{School of Artificial Intelligence and Data Science}
  \institution{University of Science and Technology of China}
  \city{Hefei}
  \country{China}}
\email{fulifeng93@gmail.com}
\authornote{Corresponding author.}

\renewcommand{\shortauthors}{Yang Zhang et al.}

\begin{abstract}
 
Recommendation unlearning is an emerging task to serve users for erasing unusable data (\emph{e.g., } some historical behaviors) from a well-trained recommender model. Existing methods process unlearning requests by fully or partially retraining the model after removing the unusable data. However, these methods are impractical due to the high computation cost of full retraining and the highly possible performance damage of partial training. In this light, a desired recommendation unlearning method should obtain a similar model as full retraining in a more efficient manner, \emph{i.e., } achieving complete, efficient and harmless unlearning.

In this work, we propose a new \textit{Influence Function-based Recommendation Unlearning} (IFRU) framework, which efficiently updates the model without retraining by estimating the influence of the unusable data on the model via the \textit{influence function}. In the light that recent recommender models use historical data for both the constructions of the optimization loss and the computational graph (\emph{e.g., } neighborhood aggregation), IFRU jointly estimates the direct influence of unusable data on optimization loss and the spillover influence on the computational graph to pursue complete unlearning. Furthermore, we propose an importance-based pruning algorithm to reduce the cost of the influence function. IFRU is harmless and applicable to mainstream differentiable models. Extensive experiments demonstrate that IFRU achieves more than 250 times acceleration compared to retraining-based methods with recommendation performance comparable to full retraining. Codes are available at \url{https://github.com/baiyimeng/IFRU}.

% Recommendation unlearning is an emerging task to serve users for erasing unusable data from a well-trained recommender model. Existing methods process unlearning requests by fully or partially retraining the model after removing the unusable data. However, these methods are impractical due to the high computation cost of full retraining and the highly possible performance damage of partial training. In this light, a desired recommendation unlearning method should obtain a similar model as full retraining in a more efficient manner, i.e., achieving complete, efficient and harmless unlearning.

% In this work, we propose a new Influence Function-based Recommendation Unlearning (IFRU) framework, which efficiently updates the model without retraining by estimating the influence of the unusable data on the model via the influence function. In the light that recent recommender models use historical data for both the constructions of the optimization loss and the computational graph, IFRU jointly estimates the direct influence of unusable data on optimization loss and the spillover influence on the computational graph to pursue complete unlearning. Furthermore, we propose an importance-based pruning algorithm to reduce the cost of the influence function. Extensive experiments demonstrate that IFRU achieves more than 250 times acceleration compared to retraining-based methods with recommendation performance comparable to full retraining.
     
\end{abstract}




\received[Received]{2 September 2023}
\received[revised]{9 July 2024}
\received[accepted]{22 September 2024}

\maketitle

\section{Introduction}

\begin{figure}[t]
  \centering
  \includegraphics[width=0.73\linewidth]{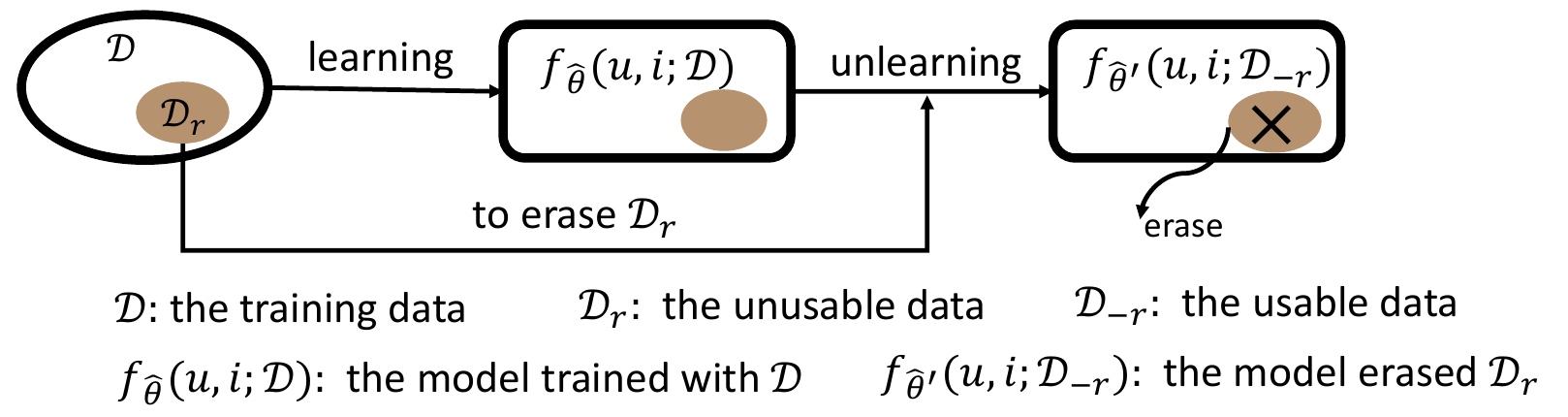}
  \caption{Illustration of recommendation unlearning process.}
  \label{fig:unlearning-flow}
  \Description{..}
\end{figure}

Recommender systems are widely deployed to perform personalized information filtering, shaping the individual experience of online activities~\cite{rec-ethical}, \eg social interactions, shopping, and entertainment. Recommender models are typically learned by fitting historical interactions, making model parameters memorize user behaviors~\cite{SML,Recunlearn,modern-recsys,deep-recsys}. The requirement to erase some training data (termed \textit{unusable data}) from trained models for some ethical and legal concerns is continuously increasing. For example, users may request to erase sensitive interaction data~\cite{Recunlearn}, and the system must take responsible responses in accordance with existing regulations such as GDPR\footnote{GDPR is short for the General Data Protection Regulation~\cite{GDPR}.}~\cite{machineUnlearn}. Besides, the system also needs to erase misinformation injected by attackers once detected to avoid poisoned recommendations related to the misinformation ~\cite{misinformation-chi,misinformation-recsys}. \textit{Recommendation unlearning} (Figure \ref{fig:unlearning-flow}) thus becomes a critical task for building trustworthy recommender systems.

Existing work achieves recommendation unlearning via retraining from scratch\footnote{Random initialization or the checkpoint before unusable data joined the training.}, which can naturally erase unusable data from the model. There are two types of retraining methods: full retraining and partial retraining~\cite{Recunlearn,LASER,yuan2022federated}. Full retraining retrains the whole model with all usable data, which is time-consuming and computation-costly~\cite{SML,Recunlearn}, prohibiting its practical use~\cite{Recunlearn}. Partial retraining accelerates unlearning via data and model partitions~\cite{Recunlearn,LASER} or gradient approximation~\cite{yuan2022federated}. However, partial retraining is typically not harmless compared to full retraining, as both the model partition and gradient approximation could hurt recommendation performance. Moreover, partial retraining may fail to achieve efficient unlearning in real-world scenarios. For example, model partition works efficiently only if unusable data is distributed locally, which is hardly satisfied in practice.

Given the pros and cons of existing methods, we distill the target of recommendation unlearning as achieving complete, efficient, and harmless unlearning. 1) Complete unlearning is a necessary condition of recommendation unlearning, \ie completely removing the influence of unusable data to make the recommender system compliant to the regulation. 2) The unlearning process should be accomplished in a sufficiently efficient manner to timely respond to unlearning requests. Otherwise, it will disrupt the normal operations of recommender systems, hurting their user experience. 3) In addition to efficiency, the unlearning process should be harmless, \ie no recommendation performance damage as compared to full retraining. In other words, we set the target of recommendation unlearning as obtaining a similar model as full retraining in a sufficiently efficient manner to timely process unlearning requests.

In this work, we investigate how to achieve recommendation unlearning without modeling retraining. \textit{Influence function}~\cite{hampel1974influence} seems to be a promising solution~\cite{li2023selective}, which is a powerful technique from robust statistics to quantify the influence of data on learning models. In this way, we can estimate the model changes caused by erasing the unusable data; and achieve efficient and harmless unlearning by updating the trained model according to the estimated changes in one step. Nevertheless, the primer influence function may fail to achieve complete unlearning since it only quantifies the direct influence of unusable data on the optimization loss, while the unusable data also affect the computational graph of recommender models. Taking the widely used LightGCN~\cite{lightgcn} as an example, erasing the unusable data (\ie removing edges from the user-item graph) will change the graph convolution operation on related nodes.

We propose a new \textit{Influence Function-based Recommendation Unlearning} (IFRU) framework to achieve more complete unlearning in recommendation. IFRU extends the influence function to quantify the influence of unusable data in the computational graph aspect, termed spillover influence. Specifically, IFRU expresses the computational graph change after erasing the unusable data in the optimization loss form, making spillover influence measurable like the direct influence. Taking one step further, we reduce the cost of influence function that computes the Hessian matrix with the size of model parameters by an importance-based pruning method to ignore less affected model parameters. IFRU avoids alterations to the training process and model architecture, emphasizing its capacity to maintain harmlessness while facilitating its application into mainstream models. We instantiate IFRU on two representative recommender models: MF~\cite{koren2009matrix} and LightGCN~\cite{lightgcn}, and conduct experiments on two real-world datasets to demonstrate our method.

The main contributions of this work are summarized as follows:
\begin{itemize}

    \item We highlight the advantages and challenges of using influence functions to achieve better unlearning in terms of completeness, efficiency, and harmlessness.

    \item We propose a new Influence Function-based Recommendation Unlearning framework, extending the influence function technique to account for the spillover influence of erasing data and accelerating the influence estimation with a new importance-based pruning strategy.
    
    \item We conduct extensive experiments on two real-world datasets, verifying the superiority of our proposal over existing recommendation unlearning methods.
   
\end{itemize}
\section{Preliminary}
In this section, we first present the problem formulation of recommendation unlearning and then briefly introduce two backbone recommender models. 

\subsection{Problem Formulation} \label{sec:def-unlearn}
\paragraph{Recommendation} 
Let $\mathcal{D}$ denote the historical interactions and $(u,i,y_{ui}) \in \mathcal{D}$ denote an interaction (\eg click) between user $u$ and item $i$, where $y_{ui}\in \{0,1\}$ denotes label of the interaction. Let $\hat{y}_{ui} = f_{\theta}(u, i; \mathcal{D})$ denote a recommender model to predict the interaction label, where $\theta$ represents the learnable parameters of the model\footnote{Note that the computational graph of a recommender model $f_{\theta}(u, i; \mathcal{D})$ could depend on the historical interactions $\mathcal{D}$.}. The conventional setting of the recommendation problem~\cite{CF-survey} is to learn the model parameters by fitting the historical data $\mathcal{D}$. Formally, the objective is defined as
\begin{equation}\label{eq:training-obj}
\begin{split}
     \hat{\theta} &= \mathop{\arg\min}\limits_{\theta} ~ L(\theta; \mathcal{D}) = \mathop{\arg\min}\limits_{\theta} ~ \frac{1}{|\mathcal{D}|} \sum_{(u,i,y_{ui}) \in \mathcal{D}} \ell \left(
     f_\theta(u,i;\mathcal{D}), y_{ui}
     \right),
\end{split}
\end{equation}
where $\hat{\theta}$ denotes the model parameters learned from $\mathcal{D}$, $L(\theta; \mathcal{D})$ denotes the total loss on $\mathcal{D}$, and $\ell(\cdot)$ is a recommendation loss function. In this study, our primary focus lies on the point-wise loss, particularly in investigating the Binary Cross-Entropy (BCE) loss. Nonetheless, our method can readily extend to other types of losses, such as pair-wise loss BPR~\cite{rendle2012bpr}.

\paragraph{Recommendation unlearning}
Let $\mathcal{D}_{r} \in \mathcal{D}$ denotes the unusable data requested by users to be erased. The target of recommendation unlearning is to erase the ``signals'' in $\mathcal{D}_{r}$ from the current recommender model $f_{\hat{\theta}}(u, i; \mathcal{D})$. In other words, the target is to answer the question: \textit{how would the model $f_{\theta}$ be if we do not have $\mathcal{D}_{r}$ during training?} Formally, the target is to approach the recommender model $f_{\hat{\theta}^*}(u,i;\mathcal{D}_{-r})$ where $\mathcal{D}_{-r} = \mathcal{D} - \mathcal{D}_r$ denotes the usable data, and $\hat{\theta}^{*}$ denotes the model parameters learned from $\mathcal{D}_{-r}$, \ie
\begin{equation}\label{eq:retraining-loss} 
 \hat{\theta}^{*} = \mathop{\arg\min}\limits_{\theta} {L(\theta, \mathcal{D}_{-r})} = \mathop{\arg\min}\limits_{\theta} \frac{1}{|\mathcal{D}_{-r}|} \sum_{(u,i,y_{ui}) \in \mathcal{D}_{-r}} \ell(f_\theta(u,i;\mathcal{D}_{-r}), y_{ui}).
\end{equation}

\paragraph{Complete-efficient-harmless Unlearning} Obviously, we can get the expected model $f_{\hat{\theta}^*}(u,i;\mathcal{D}_{-r})$ via full retraining, \ie training from scratch over the usable data $\mathcal{D}_{-r}$. Nevertheless, model retraining is time-consuming and delays the response to unlearning requests. As such, we set the target as efficiently approaching the expected model $f_{\hat{\theta}^*}(u,i;\mathcal{D}_{-r})$ according to the current model $f_{\hat{\theta}}(u,i;\mathcal{D})$ and the unusable data $\mathcal{D}_r$. Formally,
\begin{equation}
    f_{\hat{\theta}^*}(u,i;\mathcal{D}_{-r}) \stackrel{\mathcal{D}_r}{\longleftarrow} f_{\hat{\theta}}(u,i;\mathcal{D}).
\end{equation}

In our specific context, we direct our attention to a scenario where $f_{\hat{\theta}^*}(u,i;\mathcal{D}{-r})$ serves as the unaltered ground truth for unlearning. That means if $f_{\hat{\theta}}(u, i;\mathcal{D})$ can closely approximate the ground-truth model where the model parameters and computational graph align with expectations, we deem that the complete unlearning is achieved  (at least in terms of the parametric aspect). Importantly, our analysis does not encompass the attack scenarios discussed in~\cite{thudi2022necessity}, where one can manipulate the expected model to implement an approach that avoids altering the model parameters while still claiming to approximate the expected model without genuinely erasing data.

\subsection{Recommender Models}\label{sec:backbones}
Collaborative filtering (CF)~\cite{CF-survey} is one of the most representative recommendation technologies.
Without losing generality, we take two representative CF models to study recommendation unlearning:

\noindent $\bullet$ \textbf{Matrix factorization (MF)~\cite{koren2009matrix}} is one of the most classical CF models, which associates each user and item with an embedding. Let $\bm{p}_u \in \mathcal{R}^{d}$ and $\bm{q}_i \in \mathcal{R}^{d} $ denote the embedding for user $u$ and item $i$, where $d$ denotes the embedding size. MF calculates the prediction $\hat{y}_{ui}$ as the inner product of $\bm{p}_u$ and $\bm{q}_i$. Formally,
\begin{equation}
    \label{eq:mf}
    f_{\theta} (u,i;\mathcal{D}) =  \bm{p}_{u}^{\top}  \bm{q}_i,
\end{equation}
where $\theta = \{\bm{p}_u\}_{u} \cup \{ \bm{q}_i\}_{i}$, including all user and item embeddings.

\vspace{+5pt}
\noindent $\bullet$ \textbf{LightGCN~\cite{lightgcn}} is a representative graph-based CF model, which performs graph convolution over the user-item graph to aggregate user and item embeddings as follows:
\begin{equation} \label{eq:agg}
    \begin{split}
        \bm{p}_{u}^{k+1} = \sum_{i \in \mathcal{N}_{u}} \frac{1}{\sqrt{|\mathcal{N}_{u}||\mathcal{N}_{i}|}} \bm{q}_{i}^{k}; \quad
        \bm{q}_{i}^{k+1} = \sum_{u \in \mathcal{N}_{i}} \frac{1}{\sqrt{|\mathcal{N}_{i}||\mathcal{N}_{u}|}} \bm{p}_{u}^{k}
    \end{split},
\end{equation}
where $\bm{p}_{u}^{k}$ ($\bm{q}_{i}^{k}$) denotes the $k$-th layer’s representation of user $u$ (item i), $\bm{p}_{u}^{0} = \bm{p}_{u}$ ($\bm{q}_{i}^{0} = \bm{q}_{i}$), and $\mathcal{N}_{u}$ denotes the set of items positively interacted by $u$ and its size is denoted as $|\mathcal{N}_{u}|$, similarly for $\mathcal{N}_{i}$. LightGCN calculates the prediction $\hat{y}_{ui}$ from aggregated representations as follows:
\begin{equation}\label{eq:lgcn-represetation}
    f_{\theta} (u,i;\mathcal{D}) = \left(
        \sum_{k=0}^{K} \frac{1}{K+1} \bm{p}_{u}^{k}
    \right)^T \left(
        \sum_{k=0}^{K} \frac{1}{K+1} \bm{q}_{i}^{k}
    \right),
\end{equation}
where $K$ is the number of graph convolution layers; $\theta = \{\bm{p}_{u}\}_{u} \cup \{\bm{q}_{i}\}_{i} $. Note that the computational graph of LightGCN depends on $\mathcal{D}$, which is used for the construction of the user-item graph.  
In other words, the prediction $f_{\theta} (u,i;\mathcal{D})$ for some user-item pair $(u,i)$ will change if the graph is edited (\eg $\mathcal{D} \longrightarrow \mathcal{D}_{-r}$). This is because the aggregation process in Equation~\eqref{eq:agg} could be different as $\mathcal{N}_{u}$ or $\mathcal{N}_{i}$ could change.
\section{Methodology}\label{Method}
In this section, we first introduce how IFRU achieves complete-efficient-harmless recommendation unlearning via the influence function (Section~\ref{ssec:unlearn}), and further acceleration via pruning (Section~\ref{sec:IFRU}). We then present the instantiation of IFRU (Section~\ref{Instantiation}) and further discussions of the proposal (Section~\ref{ssec:discussion}).

\subsection{Recommendation Unlearning via Influence}\label{ssec:unlearn}

To achieve complete-efficient-harmless unlearning, we investigate the method to approach the expected model $f_{\hat{\theta}^*}(u,i;\mathcal{D}_{-r})$ without the time-consuming model training. The key lies in obtaining the expected model parameters $\hat{\theta}^{*}$ from the current model parameters $\hat{\theta}$. Considering that $\hat{\theta}^{*}$ and $\hat{\theta}$ corresponds to two training processes $\hat{\theta} \stackrel{\mathcal{D}}{\longleftarrow} \theta \stackrel{\mathcal{D}_{-r}}{\longrightarrow} \hat{\theta}^{*}$, we investigate how the data change ($\mathcal{D} \longrightarrow \mathcal{D}_{-r}$) influences the model parameters. In other words, we estimate the model parameter changes $\Delta = \hat{\theta}^{*} - \hat{\theta}$ caused by erasing the unusable data $\mathcal{D}_{r} = \mathcal{D} - \mathcal{D}_{-r}$; and update the trained model according to $\Delta$ in one step. In the following, we first present how to derive $\Delta$ via the influence function; and how to rapidly calculate $\Delta$.

\begin{figure}[t]
  \centering
  \includegraphics[width=0.58\linewidth]{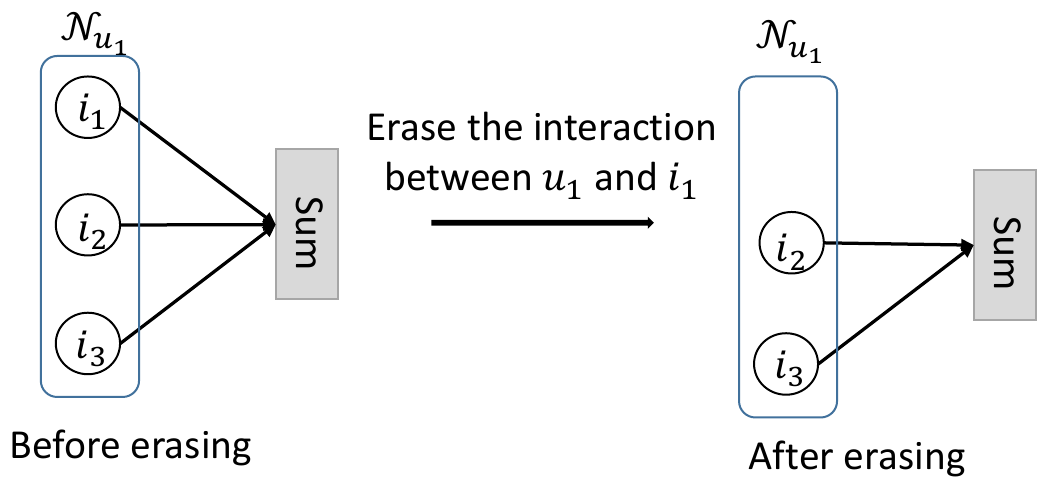}
  \caption{The neighborhood aggregation process before and after erasing an interaction in LightGCN. $\mathcal{N}_{u_1}$ represents the set of items interacted by user $u_1$.}
  \label{fig:change-example}
  \Description{..}
\end{figure}

\subsubsection{Derivation of $\Delta$}

Given a model, the learned model parameters are determined by the optimization objective. Consequently, deriving $\Delta$ hinges on scrutinizing the differences between optimization objectives with and without the unusable data $\mathcal{D}_{r}$, delineated in Equation~\eqref{eq:training-obj} and Equation~\eqref{eq:retraining-loss}. From this discrepancy, we can deduce alterations in the model parameters, thereby aiding in the acquisition of the unlearned model.

\vspace{+5pt}
\noindent \textit{\textbf{1) Impact of Data Removal on the Optimization Objective.}} In recommendation systems, eliminating unusable data points $\mathcal{D}_{r}$ not only influences the optimization term associated with those points but also has the potential to impact the optimization objective of the remaining data points by affecting the computational graph for them. For example, removing one interaction of a user will change the computational graph for other interactions of the user in LightGCN, as the aggregation process for the user (Equation~\eqref{eq:agg}) will vary. As shown in Figure~\ref{fig:change-example}, the aggregation process is affected by the neighbor set $\mathcal{N}_{u}$, while removing $u$'s interaction will change $\mathcal{N}_{u}$, leading to the change of the aggregation process. Let $\mathcal{D}_{c}$ represent the data points whose computational graph is affected. With this notation, we can reformulate the optimization objective without the unusable data $\mathcal{D}_{r}$ in Equation~\eqref{eq:retraining-loss} as follows:
\begin{equation} \label{eq:L-total}
    \begin{split}
          \mathop{\arg\min}\limits_{\theta} L(\theta;\mathcal{D}_{-r}) = \mathop{\arg\min}\limits_{\theta} \frac{|\mathcal{D}|}{|\mathcal{D}_{-r}|} \bigg[ L(\theta,\mathcal{D}) & - \frac{1}{|\mathcal{D}|} L_d(\theta;\mathcal{D}_{r}) - \frac{1}{|\mathcal{D}|} L_s(\theta;\mathcal{D}_{r})\bigg],
    \end{split}
\end{equation}
where $L_{d}(\theta;\mathcal{D}_{r})$ represents the loss term for data points $\mathcal{D}_{r}$, and $L_{s}(\theta;\mathcal{D}_{r})$ denotes the loss for data points whose computational graph has been affected by the removal of $\mathcal{D}_{r}$. In essence, $L_{d}(\theta;\mathcal{D}_{r})$ and $L{s}(\theta;\mathcal{D}_{r})$ encapsulate the total discrepancy between the optimization objectives with and without $\mathcal{D}_{r}$. Specifically, we have:
\begin{equation} \label{eq:L-direct}
L_{d}(\theta; \mathcal{D}_{r}) = \sum_{(u,i,y_{ui})\in \mathcal{D}_{r}} \ell\big(f_{\theta}(u,i;\mathcal{D}),y_{ui}\big),
\end{equation}

\begin{equation} \label{eq:L-spillover}
    \begin{split}
          L_{s}(\theta;\mathcal{D}_{r}) =  \sum_{(u,i,y_{ui})\in \mathcal{D}_{c}}  \ell\big(f_{\theta}(u,i;\mathcal{D}),y_{ui}\big) -  \ell\big(f_{\theta}(u,i;\mathcal{D}_{-r}),y_{ui}\big).
    \end{split}
\end{equation}
Here, $\mathcal{D}_c$ can be formulated as follows:
\begin{equation}\label{eq:data-c}
\mathcal{D}_{c}=\big \{(u,i,y_{u,i}) \big|(u,i,y_{u,i}) \in \mathcal{D}_{-r} \cap \exists  \theta \, f_{\theta}(u,i;\mathcal{D}) \neq f_{\theta} (u,i;\mathcal{D}_{-r}) \big \}.
\end{equation}

\vspace{+5pt}
\noindent \textit{\textbf{2) Influence of Data Removal on the Learned Model Parameters.}} 
Next, we define the influence of unusable data $\mathcal{D}_{r}$ on the model parameters using the identified loss differences. The crux lies in examining the optimal model parameters when $L_{d}(\theta;\mathcal{D}_{r})$ and $L_{s}(\theta;\mathcal{D}_{r})$ are incorporated into $L(\theta;\mathcal{D})$ with a small weight $\epsilon$, that is:
\begin{equation} \label{eq:upweighting}
    \hat{\theta}_{\epsilon} \overset{def}{=} \arg\min_{\theta} \,  L(\theta;\mathcal{D}) + \epsilon L_{d}(\theta;\mathcal{D}_{r}) + \epsilon L_{s}(\theta;\mathcal{D}_{r}).
\end{equation}
Here, $\hat{\theta}_{\epsilon}$ represents the optimal model parameters under this scenario.

It is evident that $\hat{\theta}_{0}$ ($\epsilon=0$) equals $\hat{\theta}$, and $\hat{\theta}{-\frac{1}{|\mathcal{D}|}}$ equals\footnote{When $\epsilon=-\frac{1}{|\mathcal{D}|}$, the difference between the losses in Equation~\eqref{eq:upweighting} and Equation~\eqref{eq:L-total} is the coefficient $\frac{|\mathcal{D}|}{|\mathcal{D}_{-r}|}$, which will not affect the optimal model parameters.}  $\hat{\theta}^{*}$. According to the Taylor's formula, we then have:
\begin{equation}\label{eq:taylor}
\begin{split}
    \Delta &= \hat{\theta}^{*} - \hat{\theta} =\hat{\theta}_{\epsilon=-\frac{1}{|\mathcal{D}|}} - \hat{\theta}_{\epsilon=0} \\
    &=-\frac{1}{|\mathcal{D}|} \frac{d\hat{\theta}_{\epsilon}}{d\epsilon}\big|_{\epsilon=0} + o(\frac{1}{|\mathcal{D}|}),
\end{split}
\end{equation}
where $\frac{d\hat{\theta}_{\epsilon}}{d\epsilon}\big|_{\epsilon=0}$ denotes the derivative of $\hat{\theta}_{\epsilon}$ at the point $\epsilon=0$, and $o(\frac{1}{|\mathcal{D}|})$ denotes an infinitesimal quantity which becomes arbitrarily small as the size of the dataset tends to infinity. 

With this equation, we can define the influence of erasing $\mathcal{D}_{r}$, denoted as $I(\hat{\theta};\mathcal{D}_{r})$, as $\frac{d\hat{\theta}_{\epsilon}}{d\epsilon}\big|_{\epsilon=0}$. According to classical results~\cite{IF, cook1982residuals}, we have:
\begin{equation} \label{eq:influence}
\begin{split}
        \mathcal{I} (\hat{\theta}; \mathcal{D}_{r})  \overset{def}{=}  \frac{d \hat{\theta}_{\epsilon}}{d\epsilon}\big|_{\epsilon=0}
        = \underbrace{-H^{-1}_{\hat{\theta}} \nabla_{\theta} L_{d}(\hat{\theta};\mathcal{D}_{r})}_{I_{d}(\hat{\theta};\mathcal{D}_{r})}\underbrace{
        -H^{-1}_{\hat{\theta}} \nabla_{\theta} L_{s}(\hat{\theta};\mathcal{D}_{r})}_{I_{s}(\hat{\theta};\mathcal{D}_{r})},
\end{split}
\end{equation} 

where $H_{\hat{\theta}} = \nabla^{2}_\theta L(\hat{\theta};\mathcal{D})$ is the Hessian matrix and is positive definite by assumption following previous work~\cite{IF,cheng2019incorporating}. As the equation shows, our total influence of $\mathcal{D}_{r}$ contains two parts: 1) the direct influence $\mathcal{I}_{d} (\hat{\theta};\mathcal{D}_{r})$, which is generated from the direct loss difference $L_{d}(\theta;\mathcal{D}_{r})$, representing the missing loss of $\mathcal{D}_{r}$, and 2) the spillover influence $\mathcal{I}_{s}(\hat{\theta};\mathcal{D}_{r})$, which is generated from the loss difference $L_{s}(\theta;\mathcal{D}_{r})$ caused by the change of the computational graph. We term the latter part of influence  "spillover influence" because it is generated from the changes related to the remaining data $\mathcal{D}_{c}$ instead of $\mathcal{D}_{r}$ itself, as the definition of $L_{s}(\theta;\mathcal{D}_{r})$ in Equation~\eqref{eq:L-spillover} shows.

\textit{Remarks.} The primer influence function only models the direct influence ($I_{d}(\hat{\theta};\mathcal{D}_{r})$), \ie the one caused by removing such data in the optimization loss. We further consider the spillover influence ($I_{s}(\hat{\theta};\mathcal{D}_{r})$) to capture the influence of erasing $\mathcal{D}_{r}$ on the computational graph of other remaining data, aiming at achieving more complete unlearning.

\noindent \textbf{3) Unlearned Model.} 
With the derived $\Delta$ in Equation~\eqref{eq:taylor}, we can update $\hat{\theta}$ to obtain our unlearned model parameters $\hat{\theta}^{\prime}$ as follows: 
\begin{equation}\label{eq:if-unlearn}
    \hat{\theta}^{\prime} = \hat{\theta} -\frac{1}{|\mathcal{D}|} \mathcal{I} (\hat{\theta};\mathcal{D}_{r})+o(\frac{1}{|\mathcal{D}|}) \approx \hat{\theta} -\frac{1}{|\mathcal{D}|} \mathcal{I} (\hat{\theta};\mathcal{D}_{r}).
\end{equation}
Here, we ignore $o(\frac{1}{|\mathcal{D}|})$ since $|\mathcal{D}|$ is usually large enough. Then, our unlearned model can be represented as $f_{\hat{\theta}^{\prime}}(u,i;\mathcal{D}_{-r})$. The unlearning can be achieved by one step after getting $\mathcal{I} (\hat{\theta};\mathcal{D}_{r})$. We next consider how to calculate $\mathcal{I} (\hat{\theta};\mathcal{D}_{r})$.

\begin{figure}[t]
  \centering
  \includegraphics[width=0.55\linewidth]{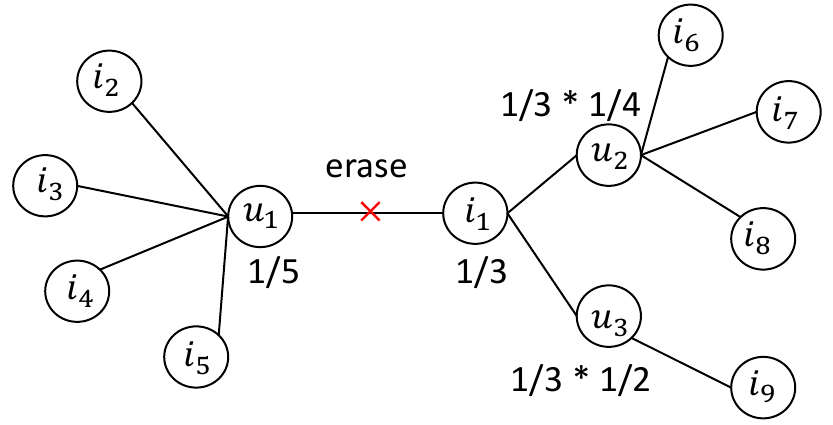}
  \caption{Illustration of how erasing an interaction between $u_1$ and $i_1$ affects users and items. The value around a node denotes its importance score.
  }
  \label{fig:pruning}
  \Description{..}
\end{figure}

\subsubsection{Calculating the Influence $I(\hat{\theta};\mathcal{D}_{r})$.} \label{sec:calculating}
 
Directly calculating $I(\hat{\theta};\mathcal{D}_{r})$ could be still computation-costly since it involves the calculation of $H_{\hat{\theta}}$ and $H^{-1}_{\hat{\theta}}$. Let $n$ denote the size of training data ($n=|\mathcal{D}|$) and $p$ denote the size of model parameters. The total computation cost is $O(np^2 + p^3)$, since directly computing $H_{\hat{\theta}}$ and $H^{-1}_{\hat{\theta}}$ need $O(np^2)$ and $O(p^3)$ operations, respectively. Fortunately, we can also take \textit{Conjugate Gradients} (CG) and \textit{Stochastic Estimation} (SE) methods proposed in~\cite{IF} to reduce the complexity to $O(np)$. 
 
In this work, we modify the CG method to make the solution suitable for existing machine learning frameworks like PyTorch. In short, we convert the problem of computing $I(\hat{\theta},\mathcal{D}_{r})$ into an optimization problem like the CG method but solve it with the Adam optimizer~\cite{adam}. Specially, let $t^{*}=-\nabla_{\theta}L_{d}(\hat{\theta};\mathcal{D}_{r}) - \nabla_{\theta}L_{s}(\hat{\theta};\mathcal{D}_{r}) $, and then it is easily verified~\cite{IF} that\footnote{With the assumption that $H_{\hat{\theta}}$ is positive definite.}:
 \begin{equation}\label{eq:optimize-influence}
    I(\hat{\theta},\mathcal{D}_{r})=H_{\hat{\theta}}^{-1} t^{*} = argmin_{t}\, \frac{1}{2}t^{\top}H_{\hat{\theta}}t-t^{\top}t^{*},
 \end{equation}
where $t\in\mathbb{R}^{p}$ is a learnable vector. We can calculate the influence by minimizing the loss on the right of the equation. For the loss,  the gradient of $t$ is always $H_{\hat{\theta}}t-t^{*}$. Then, we only need to input the gradient into the Adam optimizer to update $t$. When computing the gradient, to avoid explicitly computing $H_{\hat{\theta}}$, which is space-costly, we take the \textit{Hessian Vector Product} (HVP) to efficiently compute $H_{\hat{\theta}}t$ as follows: 
\begin{equation}\label{eq:hvp}
    H_{\hat{\theta}}t = \nabla^{2}_{\theta}L(\hat{\theta};\mathcal{D})t =  \nabla_{\theta}(\nabla_{\theta}L(\hat{\theta};\mathcal{D})^{\top}t),
\end{equation}
the computation cost of which is $O(np)$. As we could get a good result for this second-order optimization problem in a few iterations with the Adam optimizer, the computational cost of this method can be near $O(np)$.
 
\subsection{Unlearning with Pruning} \label{sec:IFRU}
In CF models, each user (item) usually corresponds to special parameters, \eg an embedding.  That means the size of model parameters (\ie $p$ in $O(np)$) could be very large, limiting the response speed of our method to the real-time unlearning request. We thus further consider pruning some unimportant model parameters, which are less affected by erasing $\mathcal{D}_{r}$, to make $p$ smaller.

\vspace{+5pt}
\noindent \textbf{Pruning model parameter.} 
As user and item embeddings dominate the model parameters in CF models, we focus on pruning them. 
Considering erasing an interaction, we find both the user and the item themselves and their neighbors will be affected. Figure~\ref{fig:pruning} shows an example:  erasing the interaction between $u_{1}$ and $i_{1}$ will 1) directly affect $i_{1}$, and 2) indirectly affect $u_{2}$ and $u_{3}$, the 1-order neighbors of $i_{1}$, since $i_{1}$ and $u_{2}$ ($u_{3}$) affect learning each other when fitting the interaction between them. These users and items are undoubtedly affected to different degrees. We develop a general method to analyze the extent to which users/items (embeddings) are affected by erasing $\mathcal{D}_{r}$ from the fitting perspective without considering the model architecture. Algorithm~\ref{alg:pruning} summarizes our method, which contains two parts to analyze the directly affected users/items ($0$-order neighbors) and indirectly affected users/items (high order neighbors), respectively:

\begin{itemize}[leftmargin=*]
    \item[-] \textbf{Part 1} (lines 3-7) considers the users and items of $\mathcal{D}_{r}$, which are directly affected by erasing $\mathcal{D}_{r}$. As each interaction contributes equally to the training loss in form, we roughly assume that each interaction equally contributes to learning the user/item embedding. We then naturally define $\frac{1}{|\mathcal{N}_{u}|}$ as the degree to which $u$ is affected by erasing a $u$'s interaction, similarly for $i$, named importance score. For the example in Figure~\ref{fig:pruning}, the importance score gotten by $u_1$ ($i_1$) is $1/5$ ($1/3$). Let $v$ denote a user or an item. For each $v$, the importance scores caused by erasing different interactions are accumulated to form the final importance score, denoted as $s^{0}_{v}$. We only keep the users and items that have higher final important scores with the ratio of $a_{0}\in [0,1.0]$ and prune others, which are less affected.

    \item[-] \textbf{Part 2} (lines 9-16) considers the $k$-order neighbors ($k>=1$). After getting the importance score of $(k-1)$-order neighbor $v$ (denoted as $s_{v}^{k-1}$), we define the importance score gotten by $k$-order neighbor $v^{\prime}$ (interacted with $v$) is $s_{v}^{k-1} * \frac{1}{|\mathcal{N}_{v^{\prime}}|}$. This is because we similarly assume the neighbors of $v^{\prime}$ affect $v^{\prime}$ in the same degree. For example in Figure~\ref{fig:pruning},  the importance score gotten by the $1$-order neighbor $u_{2}$ from $0$-order neighbor $i_{1}$ is $1/3*1/4$. We accumulate all importance scores to form the total importance score $s_{v^\prime}^{k}$ for the $k$-order neighbor $v^\prime$.
    We then similarly prune users and items with a ratio of $a_{k}$ $\in [0.0, 1.0]$ according to the total importance score.    
\end{itemize}

With these pruning operations, the algorithm will retain the important users and items for unlearning and return their model parameters $\phi$ (lines 17).

\begin{algorithm}[t]
	\caption{Pruning}
	\LinesNumbered
	\label{alg:pruning}
	\KwIn{ Pruning ratios $\{a_0,...,a_K\}$, hyper-parameters $K$, training data $\mathcal{D}$, and the unusable data $\mathcal{D}_{r}$}. 
    For each $k\in\{0,1,\dots,K\}$, initialize $s_v^{k}$ as zero for each $v$ (a user or an item), and initialize $\mathcal{S}^{k}=\varnothing$ \;
    // Part 1 ($0$-order neighbors)\;
    \For{$(u,i,y_{ui}) \in \mathcal{D}_{r} $}{
    $s_{v}^{0} \leftarrow  s_{v}^{0} + \frac{1}{|\mathcal{N}_{v}|}$ for $v \in \{u,i\}$\; 
    Add $ \{u,i\}$ into $\mathcal{S}^{0}$\;
    }
    Update $\mathcal{S}^{0} \leftarrow \{v| v \in \text{the top}\, a_{0} \, \text{of}\, \mathcal{S}^{0} \wrt s_{v}^{0}\}$ \;
    // Part 2 ($k$-order neighbors)\;
    \For{$k = 1,\dots,K $}{
    Let $s_{v}^{k} \leftarrow s_{v}^{k-1}$ for all $v$\;
    \For{ each $ v \in S^{k-1} $}{
    Update $s_{v^{\prime}}^{k} \leftarrow s_{v^{\prime}}^{k} + s_{v}^{k-1} *  \frac{1}{\mathcal{N}_{v^{\prime}}} $  for all $v^{\prime} \in \mathcal{N}_{v} $\;
    Add $\mathcal{N}_{v}$ into $\mathcal{S}^{k}$\;
    }
    Update $\mathcal{S}^{k} \leftarrow \{v| v \in \text{the top}\, a_{k} \, \text{of}\, \mathcal{S}^{k} \wrt s_{v}^{k}\}$ \;
    }
    return $\phi$ = \{model parameters of $v$ | $v\in  S^{0}\cup S^{1}\dots\cup S^{K}$\}\;
\end{algorithm}

\vspace{+5pt}
\noindent \textbf{Unlearning after pruning.} After pruning, we obtain the $\phi$ important for unlearning\footnote{Besides embeddings, there are also possible model parameters shared by all users (items) in some CF methods, \eg NCF~\cite{ncf}. They could also be added into $\phi$.}. Let $\psi$ denote the model parameters except for $\phi$, we then have $\theta = [\psi,\phi]$ and  $\hat{\theta} = [\hat{\psi},\hat{\phi}]$. To speed up the unlearning, we ignore $\psi$ and  only consider the influence of $\mathcal{D}_{r}$ on $\phi$. Towards the goal, we replace $\theta$ with $[\psi,\phi]$ and fix $\psi=\hat{\psi}$ in Equation~\eqref{eq:upweighting} to define a $\hat{\phi}_{\epsilon}$ as follows:
\begin{equation} 
\begin{split}
        \hat{\phi}_{\epsilon} \overset{def}{=}  argmin_{\phi} \,  L(\hat{\psi},\phi;\mathcal{D}) + \epsilon L_{d}(\hat{\psi},\phi;\mathcal{D}_{r}) 
    + \epsilon L_{s}(\hat{\psi},\phi;\mathcal{D}_{r})).
\end{split}
\end{equation}
Following the definition in Equation~\eqref{eq:influence}, we define the influence of $\mathcal{D}_{r}$ on $\phi$ as follows:
\begin{equation}
\begin{split}
        \mathcal{I} (\hat{\phi};\mathcal{D}_{r}) \overset{def}{=} \frac{d \hat{\phi}_{\epsilon}}{d\epsilon}\big|_{\epsilon=0} 
         = \underbrace{-H^{-1}_{\hat{\phi}} \nabla_{\phi} L_{d}(\hat{\psi},\hat{\phi};\mathcal{D}_{r})}_{\mathcal{I}_{d} (\hat{\phi};\mathcal{D}_{r})}
        \underbrace{-H^{-1}_{\hat{\phi}} \nabla_{\phi} L_{s}(\hat{\psi},\hat{\phi};\mathcal{D}_{r})}_{\mathcal{I}_{s}(\hat{\phi};\mathcal{D}_{r})},
\end{split}
\label{eq:influence-pruning}
\end{equation}
where $H^{-1}_{\hat{\phi}}=\nabla_{\phi}^{2}L(\hat{\psi},\hat{\phi};\mathcal{D})$, and $L(\hat{\psi},\hat{\phi};\mathcal{D})$ has the same meaning to $L(\hat{\theta};\mathcal{D})$, similarly for others. Next, similar to Equation~\eqref{eq:if-unlearn}, the unlearned model parameters can be obtained as follows:
\begin{equation}
    \hat{\theta}^{\prime} = [\hat{\psi}, \hat{\phi} - \frac{1}{|\mathcal{D}|} \mathcal{I} (\hat{\phi};\mathcal{D}_{r})].
\end{equation}

After pruning, we only need to estimate $\mathcal{I} (\hat{\phi};\mathcal{D}_{r})$ instead of $\mathcal{I} (\hat{\theta}; D_{r})$. Assuming the size of $\phi$ is $p^{\prime}$ ($p^{\prime}<p$), the time complexity can decrease from $O(np)$  to $ O(n^{\prime} p^{\prime})$, where $n^{\prime}$ ($n^{\prime}<n$) is the number of training points that affect computing $H_{\hat{\phi}}$.

\subsection{Instantiation}\label{Instantiation}

To demonstrate how our proposed IFRU works, we provide two implementations based on MF and LightGCN. The key lies in identifying $L_{d}(\theta;\mathcal{D}_{r})$ and $L_{s}(\theta;\mathcal{D}_{r})$ for computing $\mathcal{I}(\hat{\theta};\mathcal{D}_{r})$ (or $\mathcal{I}(\hat{\phi};\mathcal{D}_{r})$). $L_{d}(\theta;\mathcal{D}_{r})$ is the training loss on $\mathcal{D}_{r}$ and can be directly obtained. The key of computing $L_{s}(\hat{\theta};\mathcal{D}_{r})$ is to identify  $\mathcal{D}_{c}$, as shown in Equation~\eqref{eq:L-spillover}. We thus give $\mathcal{D}_{c}$ for MF and LightGCN.

\vspace{+5pt}
\noindent \textbf{Instantiation on MF.}  
As MF directly takes the product of the user embedding and item embedding as the prediction, erasing $\mathcal{D}_{r}$ will not change the computational graph for other remaining data, which means that: $$f_{\theta}(u,i;\mathcal{D}) = f_{\theta}(u,i;\mathcal{D}_{-r}),\, \text{for any}\, (u,i,y_{u,i})\in \mathcal{D}_{-r}.$$ Therefore, according to the definition of $\mathcal{D}_{c}$ in Equation~\eqref{eq:data-c}, $\mathcal{D}_{c}=\varnothing$, and we have: $$\mathcal{I}(\hat{\theta};\mathcal{D}_{r}) = \mathcal{I}_{d}(\hat{\theta};\mathcal{D}_{r}).$$

\vspace{+5pt}
\noindent \textbf{Instantiation on LightGCN.}  For LightGCN, the computational graph for some remaining data will possibly change if removing $\mathcal{D}_{r}$, \eg the aggregation operations could be different under $\mathcal{D}$ and  $\mathcal{D}_{-r}$ as shown in Figure~\ref{eq:agg}. That means, for some training data $(u,i,y_{ui}) \in \mathcal{D}_{-r}$, we have $f_\theta(u,i;\mathcal{D}) \neq f_\theta(u,i;\mathcal{D}_{-r})$.  Thus $\mathcal{D}_{c}$ is not $\varnothing$. For simplicity, we set the number of graph convolution layers as $1$. In this case, it is easily verified that only the interaction (in $\mathcal{D}_{-r}$) of the users/items having interactions in $\mathcal{D}_{r}$ belongs to $\mathcal{D}_{c}$. That means:
\begin{equation*}
        \mathcal{D}_{c} =\big\{(u,i,y_{ui})| (u,i,y_{ui}) \in \mathcal{D}_{-r}, \, and
         \,u \in \mathcal{V}(\mathcal{D}_{r})\, or \, i \in \mathcal{V}(\mathcal{D}_{r}) \big\},
\end{equation*}
where $\mathcal{V}(\mathcal{D}_{r})$ is the set of users and items that contain at least one interaction in $\mathcal{D}_{r}$. We have $$\mathcal{I}(\hat{\theta};\mathcal{D}_{r}) = \mathcal{I}_{d}(\hat{\theta};\mathcal{D}_{r}) + \mathcal{I}_{s}(\hat{\theta};\mathcal{D}_{r}).$$

\subsection{Discussion}\label{ssec:discussion} 
We next discuss whether our method IFRU achieves the three goals of recommendation unlearning.
\begin{itemize}[leftmargin=*]
    \item \textbf{Complete unlearning}. 
    According to Equation~\eqref{eq:taylor} and {Equation~\eqref{eq:if-unlearn}}, we can represent the retraining model parameters $\hat{\theta}^{*}$ with our unlearned model parameters $\hat{\theta}^{\prime}$ as follows:
    \begin{equation}
    \begin{split}
        \hat{\theta}^{*} 
        & = \hat{\theta} -\frac{1}{|\mathcal{D}|} \frac{d\hat{\theta}_{\epsilon}}{d\epsilon}\big|_{\epsilon=-\frac{1}{|\mathcal{D}|}} + o(\frac{1}{|\mathcal{D}|})\\
       & = \hat{\theta} -\frac{1}{|\mathcal{D}|} \mathcal{I} (\hat{\theta};\mathcal{D}_{r}) + o(\frac{1}{|\mathcal{D}|})
        =\hat{\theta}^{\prime} + o(\frac{1}{|\mathcal{D}|}),
    \end{split}
    \end{equation}
    As we consider both the spillover and direct influence, the estimated $\mathcal{I} (\hat{\theta};\mathcal{D}_{r})$ can faithfully reflect the true $\frac{d\hat{\theta}_{\epsilon}}{d\epsilon}\big|_{\epsilon=-\frac{1}{\mathcal{D}}}$. Then, if $|\mathcal{D}|$ is large enough, $o(\frac{1}{|\mathcal{D}|})$ is ignorable and $\hat{\theta}^{\prime}$ is a good approximation for the $\hat{\theta}^{*}$. Then, our unlearned model $f_{\hat{\theta}^{\prime}}(u,i;\mathcal{D}_{-r})$ can approach $f_{\hat{\theta}^{*}}(u,i;\mathcal{D}_{-r})$ well\footnote{It is important to note that there are two underlying assumptions: 1) we assume that the errors caused by randomness and convergence in the learning models and calculating the influence can be ignored, and 2) it is assumed that the objective is strongly convex. Recent research by~\cite{bae2022if} suggests that the errors caused by randomness and convergence are very small. Furthermore, under non-convex cases, we can add the identity matrix with a damping term~\cite{IF} to refine the defined influence, making it still applicable with smaller approximation errors~\cite{bae2022if,inf-unlearn}.}. Therefore, IFRU could achieve approximatively complete unlearning (defined in Section~\ref{sec:def-unlearn}) if the size of training data is large enough. Regarding the pruning, empirical results {(\cf Section~\ref{sec:pruning-exp})} show that it will only slightly sacrifice the unlearning completeness since we only prune the model parameters less affected by erasing $\mathcal{D}_{r}$.

    \item \textbf{Efficient unlearning}.
    The time complexity of our method is $O(np)$ and can further decrease to $O(n^{\prime}p^{\prime})$ with pruning. Empirical results show that our method is far faster than retraining-based methods regardless of how non-locally the unusable data $\mathcal{D}_{r}$ distributes (\eg completely random sampling) and how large the size of $\mathcal{D}_{R}$ is (\eg up to 8\% of $\mathcal{D}$ in Section~\ref{sec:time-exp}). Besides, if the backbone model {has} massive parameters, we could prune redundant parameters to keep the efficiency. If the backbone model has very few model parameters, we could even pre-compute the $H_{\theta}^{-1}$ to achieve more efficient unlearning.
    
    \item \textbf{Harmless unlearning}. 
    Our method is a full post-processing method without modifications to model and training architectures. From this perspective, \ie without considering the approximation errors of the completeness aspect, it could achieve unlearning without side effects on recommendation performance like full retraining and thus has superiority in achieving harmless unlearning, compared to existing partial retraining-based methods.

\end{itemize}
\section{Experiment}
In this section, we conduct experiments to answer two research questions:

\begin{itemize}
    \item \textbf{RQ1}: How does the proposed IFRU perform in terms of achieving the three goals of recommendation unlearning, as compared with the state-of-the-art unlearning methods?

    \item \textbf{RQ2}: How do our designs and unlearning settings affect the effectiveness and efficiency of the proposed IFRU?
\end{itemize}

\subsection{Experimental Settings}
\subsubsection{Datasets}
We conduct experiments on two benchmark datasets: Amazon-electronics and BookCrossing, which are both publicly accessible. The statistics of the datasets are summarized in Table~\ref{tab:dataset_stats}.

\noindent\textbf{Amazon-electronics}. This dataset refers to the `Electronics' subset of Amazon-review~\cite{amazon-data}, a widely used product recommendation dataset. It contains user ratings for electronic products in the range of $1\text{-}5$. We convert the ratings to binary labels (`0' or `1') with the threshold of $4$. That means we label the rating value larger than four as positive (`1') and otherwise negative (`0') and do not consider the non-rated items. Besides, we take 5-core filtering to ensure that each user (item) has at least five interactions. For simplicity, we take "Amazon" to denote the dataset later.

\noindent\textbf{BookCrossing}. This is a book rating dataset\footnote{\url{http://www2.informatik.uni-freiburg.de/~cziegler/BX}} collected from the Book-Crossing community. Similarly, the 5-core filtering is applied for the sake of data quality. We also convert the ratings (in the range of $0\text{-}10$) into binary labels. To be more specific, if the original rating is greater than $6$, we label it to `1', otherwise `0'.

We randomly partition each dataset into training, validation, and testing sets, allocating them in a ratio of 6:2:2 based on interactions.

\begin{table}[t]
    \caption{Statistical details of the evaluation datasets.}
    \label{tab:dataset_stats}
    \setlength{\tabcolsep}{1mm}
    \begin{tabular}{lllll}
    \hline
    \textbf{Dataset}      & \textbf{\#User}\ & \textbf{\#Item}\ & \textbf{\#Interaction}\ & \textbf{Density} \\ \hline
    \textbf{Amazon}       & 4,201,696       &     476,001    & 7,824,481  & 0.0003\%        \\
    \textbf{BookCrossing} &105,283 &340,556 &  1,149,780    &  0.003\% \\ \hline
    \end{tabular}
\end{table}

\subsubsection{Evaluation Setting} 
There are three goals of recommendation unlearning, \ie complete unlearning, efficient unlearning, and harmless unlearning. We can directly evaluate the unlearning efficiency by measuring the running time. To evaluate whether a method achieves complete unlearning and harmless unlearning, we take a naive solution to compare the recommendation accuracy between the backbone model retrained from scratch and the estimated unlearned model. The performance gap between the two models would indicate the failure in achieving complete unlearning or harmless unlearning or both of them. However, the solution could be unreliable if we directly erase clean training data. This is because erasing clean data would lead to performance decreases, while the decreases could also be achieved by many random factors\footnote{E.g., random perturbations in model parameters.}. That means a method could show similar performance to the backbone model retrained from scratch just by random factors.

To alleviate the issue, we take an attack setting instead of erasing clean data, similar to \cite{yuan2022federated}. Specifically, we randomly select a subset of training data and attack it by reversing data labels (\ie converting `1' to `0' and vice versa). We merge the attack data and the remaining data to form training data $\mathcal{D}$ and train a recommender model with $\mathcal{D}$. We then treat the attacked data as the unusable data $\mathcal{D}_{r}$ to erase from the trained model. Different from erasing clean training data, erasing the attacked data $\mathcal{D}_{r}$ would result in performance increases, as $\mathcal{D}_{r}$ is harmful to model learning. Performance increases are more {difficultly} achieved by random factors, and we thus believe that the attack setting makes the evaluation solution more reliable.

\textit{Metrics.} We measure the performance of accuracy with three AUC~\cite{fawcett2006introduction} (Area under the ROC Curve) based metrics: 1) $AUC_{0}$, which is the AUC computed on all testing data, 2) $AUC_{1}$, which is the AUC computed on the testing data in which the user or the item of a sample belongs to $\mathcal{V}(\mathcal{D}_{r})$ (\ie the set of users and items that contain interactions in $\mathcal{D}_{r}$), and 3) $AUC_2$, which is the AUC computed on the testing data in which both the user and item of a sample belong to $\mathcal{V}(\mathcal{D}_{r})$. $AUC_{0}$ measures global model performance while $AUC_1$ and $AUC_2$ measure the local performance of the part more affected by erasing $\mathcal{D}_{r}$.

\subsubsection{Compared Methods} 

We compare the proposed IFRU method with the following recommendation unlearning methods:
\begin{itemize}[leftmargin=*]
    \item \textbf{Retrain}. It refers to the full retraining method, which directly retrains the base mode with $\mathcal{D}_{-r}$ $ (=\mathcal{D}-\mathcal{D}_{r})$ from scratch.
    
    \item \textbf{RecEraser}~\cite{Recunlearn}. This is a partial retraining-based method. It first divides the training set into several shards using a balanced partition algorithm designed from the user aspect, item aspect, or interaction aspect based on the pre-trained embeddings. Then, it trains sub-models for different shards, followed by an adaptive aggregation module to obtain the final prediction.
    
    \item \textbf{SISA}~\cite{machineUnlearn}. This is also a partial retraining-based method. It works similarly to RecEraser, and the main difference lies in that it randomly splits training data into shards. It is not originally designed for recommendation, but we apply it to recommendation, following the previous work~\cite{Recunlearn}.

    \item \textbf{SCIF}~\cite{li2023selective}. 
    This method for unlearning is specifically designed for recommendation. It also utilizes the influence function to eliminate the impact of unusable data, but it overlooks the potential spillover effects. Additionally, rather than truly removing unusable data, it substitutes the data labels with an averaged one, possibly leading to adverse outcomes.
\end{itemize}

We implement all the compared methods on two classical collaborative filtering models: MF and LightGCN, which have been introduced in Section~\ref{sec:backbones}. Besides the compared recommendation unlearning methods, there is another partial retraining-based recommendation unlearning method named FRU~\cite{yuan2022federated}, which reduces the size of retraining data with rough gradient approximation to accelerate unlearning. We do not compare it since it is specially designed for the federated recommendation.

\subsubsection{Hyper-parameters Setting}
For fair comparisons, we use BCE loss~\cite{ncf} to learn model parameters for all methods. Meanwhile, we optimize all models using the Adam optimizer with a default batch size of 2048 and set the maximal number of epochs to 5000. The best hyper-parameters are determined based on the result of the validation set. The embeddings of the backbone recommender are initialized with the Gaussian distribution, the mean of which is fixed to zero and the standard deviation of which is tuned in the range of \{1e-2, 1e-3, 1e-4\}. The embedding size is tuned in the range of \{32, 48, 64\}. The early stopping strategy is applied to prevent the over-fitting problem, that is, terminating model training if the $AUC_0$ metric on the validation set does not increase in 50 successive epochs. The ratio of unusable data $\mathcal{D}_{r}$ to training data $\mathcal{D}$, \ie $|\mathcal{D}_{r}|/|\mathcal{D}|$ is set to $2\%$ for MF and $1\%$ for LightGCN by default\footnote{We would include a study that does not use HVP, where unlearning for LightGCN would be costly. Thus we take a smaller ratio for LightGCN to ensure the study could be easily conducted.}. For all the compared methods, we tune the learning rate in the range of \{1e-2, 1e-3, 1e-4\}. For RecEraser, we adopt the Interaction-based Balanced Partition algorithm for data splitting and set the number of shards to 10, as suggested by the authors in their paper. The number of shards in SISA is also set to 10. For IFRU, we fix $K$ in Algorithm~\ref{alg:pruning} to 1, as $1$-hop neighbors cover almost all users and items under the default ratio of unusable data. Then, we set the pruning ratios $(a_0,a_1)$ of Algorithm~\ref{alg:pruning} to (1.0, 1.0) for MF and (1.0, 0.6) for LightGCN by default. The learning rate of Adam optimizer to calculate the defined influence is tuned in the range of \{1e3, 2e3, 5e3, 1e4, 2e4, 5e4\}. Notably, we have also tried smaller values for this learning rate and larger values for other learning rates. However, both of them lead to much worse results.  

\begin{table*}[t]
    \caption{
    Recommendation accuracy of compared methods to erase attack data $\mathcal{D}_{r}$. "Original" denotes the backbone model without erasing $\mathcal{D}_{r}$, as a reference here. %\textcolor{blue}
    }
    \label{tab:overall_efficacy}
    \resizebox{1.0\textwidth}{!}{
    \begin{tabular}{ccccccclcccccc}
    \hline
    \multirow{2}{*}{Amazon}       & \multicolumn{6}{c}{MF}                                    &  & \multicolumn{6}{c}{LightGCN}                              \\ \cline{2-7} \cline{9-14} 
                                  & Original & Retrain & SISA   & RecEraser & SCIF   & IFRU   &  & Original & Retrain & SISA   & RecEraser & SCIF   & IFRU   \\ \hline
    $AUC_{0}$                     & 0.6192   & 0.6266  & 0.5447 & 0.5885    & 0.6239 & 0.6263 &  & 0.6498   & 0.6524  & 0.5807 & 0.5911    & 0.6516 & 0.6524 \\
    $AUC_{1}$                     & 0.6314   & 0.6416  & 0.5572 & 0.6015    & 0.6375 & 0.6403 &  & 0.6669   & 0.6708  & 0.5984 & 0.6075    & 0.6682 & 0.6706 \\
    $AUC_{2}$                     & 0.6359   & 0.6674  & 0.5553 & 0.6146    & 0.6654 & 0.6608 &  & 0.6661   & 0.6857  & 0.6008 & 0.6109    & 0.6859 & 0.6888 \\ \hline
    \multirow{2}{*}{BookCrossing} & \multicolumn{6}{c}{MF}                                    &  & \multicolumn{6}{c}{LightGCN}                              \\ \cline{2-7} \cline{9-14} 
                                  & Original & Retrain & SISA   & RecEraser & SCIF   & IFRU   &  & Original & Retrain & SISA   & RecEraser & SCIF   & IFRU   \\ \hline
    $AUC_{0}$                     & 0.7342   & 0.7382  & 0.7009 & 0.7315    & 0.7380 & 0.7384 &  & 0.6575   & 0.6632  & 0.6767 & 0.7091    & 0.6602 & 0.6628 \\
    $AUC_{1}$                     & 0.7407   & 0.7458  & 0.7179 & 0.7408    & 0.7453 & 0.7458 &  & 0.6663   & 0.6753  & 0.6988 & 0.7252    & 0.6686 & 0.6744 \\
    $AUC_{2}$                     & 0.7376   & 0.7536  & 0.7076 & 0.7372    & 0.7523 & 0.7513 &  & 0.6699   & 0.6989  & 0.6629 & 0.7062    & 0.6700 & 0.6933 \\ \hline
    \end{tabular}
    }
\end{table*}

\subsection{Performance Comparison (RQ1)}
\subsubsection{Accuracy Comparison}\label{sec: acc-comp}

We start by comparing the recommendation accuracy of IFRU and baselines on different backbones, to evaluate how these models achieve the goals of complete unlearning and harmless unlearning. We also report the result of `Original' that refers to the backbone model trained on $\mathcal{D}$, \ie the result before unlearning. The comparison results are summarized in Table~\ref{tab:overall_efficacy}, where we have the following observations:

\begin{figure*}
    \centering
    \includegraphics[width=0.95\linewidth]{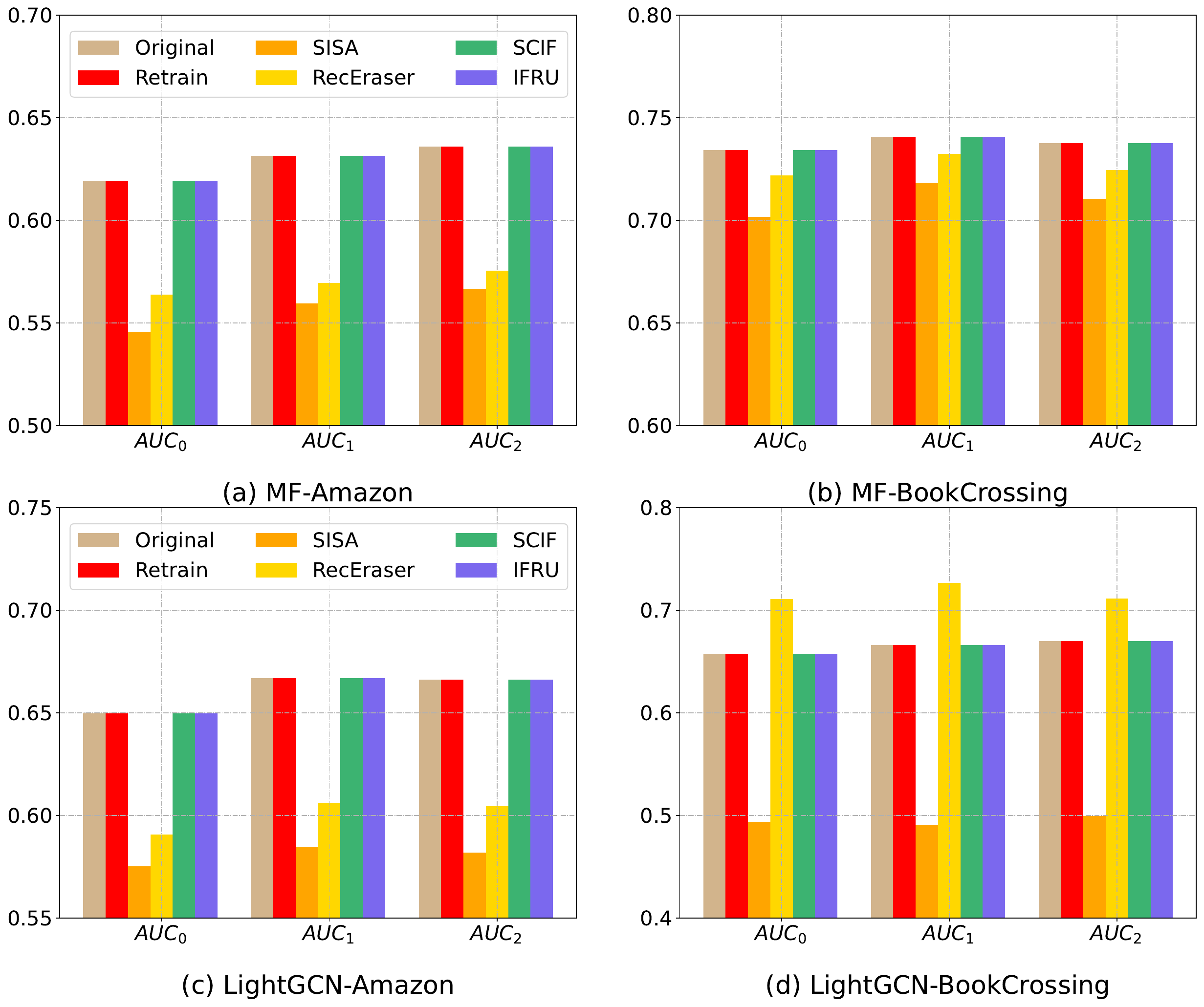}
    \caption{
    Comparison of recommendation accuracy among methods for erasing an empty data set from a trained model. The reference method "Original" denotes the backbone model without performing unlearning.
    }
    \label{fig:wuhai}
    \Description{..}
\end{figure*}

\begin{itemize}%[leftmargin=*]
  
    \item Retrain shows better performance than Original on all accuracy metrics. This verifies that in our attack setting, erasing attack data $\mathcal{D}_{r}$ should bring performance increases. Meanwhile, the performance gap between Retrain and Original is larger on $AUC_{1}$ and  $AUC_{2}$ than $AUC_{0}$, showing that users (items) are affected by erasing $\mathcal{D}_{r}$ to different degrees.

    \item IFRU achieves performance increases compared to Original, and the performance increase achieved by IFRU is very close to that of Retrain on both MF and LightGCN. Note that Retrain is the ground truth for unlearning. The results show that IFRU can erase the information of $\mathcal{D}_{r}$ from the trained backbones, and its unlearning is relatively complete. We define a completeness coefficient, the ratio of the performance increase of an unlearning method to Retrain's performance increase, to quantify the completeness of unlearning. A coefficient closer to $100\%$ indicates more complete unlearning. On BookCrossing (Amazon), the average coefficient of IFRU is $97\%$ ($88\%$)  for MF and $88\%$ ($103\%$)  for LighGCN over the three metrics\footnote{If only focusing on $AUC_{0}$, the results are very closer to $100\%$ ($99\%$ for BookCrossing, $98\%$ for Amazon). }.

    \item Regarding RecEraser and SISA, they could show better performance than Retrain on the LightGCN for BookCrossing but show worse performance even than Original in all other cases. 
    The results can be attributed to their modifications to the model and training architectures of backbones, which possibly prevent achieving harmless unlearning, \ie bringing side effects on model performance.  In addition, RecEraser shows higher accuracy than SISA, which can be attributed to the non-random data splitting of RecEraser~\cite{Recunlearn}.
    
    \item Compared to RecEraser and SISA, the performance of IFRU  is close to the backbone model (Original and Retrain), showing the superiority of achieving harmless unlearning by avoiding modifying the model and training architectures. 
    
    \item SCIF generally falls short when compared to IFRU. This can be attributed to SCIF's failure to account for spillover effects and its potential introduction of adverse effects by replacing the label of unusable data with the averaged one.
    
\end{itemize}

\textit{Harmless unlearning.} 
Both incomplete unlearning and not achieving harmless unlearning can result in performance gaps between the Retrain method and unlearning methods, which may raise doubts about the superiority of IFRU in achieving harmless unlearning. To provide direct evidence for the superiority, we conduct an additional experiment. 
Specifically, after the backbone model was trained with a dataset, we forcibly set the unusable data $\mathcal{D}_{r}$ with an empty set and performed unlearning on it. 
In this case, if there is still a performance gap to Retrain, it must be due to not achieving harmless unlearning, as removing an empty set would not bring parameter changes for any method, including IFRU (seeing Equation~\eqref{eq:influence}). The results of the experiment are summarized\footnote{Notably, the training datasets for MF and LightGCN are $\mathcal{D}$ and $\mathcal{D}_{-r}$, respectively. Utilizing $\mathcal{D}_{-r}$ leads to performance discrepancies between Figure 4 and Table 2 for LightGCN.} in Figure~\ref{fig:wuhai}.
The figure shows that IFRU/SCIF achieves the same performance as Original and Retrain, confirming that unlearning with influence function can help achieve harmless unlearning. In contrast, RecEraser and SISA show performance gaps to Original and Retrain, indicating that harmless unlearning is not achieved.
 
\begin{table*}[t]
    \caption{
    Running time (second[s]) of Retrain, RecEraser, and IFRU to erase unusable data of different sizes (BookCrossing). ‘Ratio’ denotes the ratio of unusable data $\mathcal{D}_{r}$ to $\mathcal{D}$, and a higher ratio means erasing more data. The time cost evaluation is performed on a machine with NVIDIA RTX 3090 GPUs.
    }
    \label{tab:exp_overall_efficiency}
    \resizebox{0.9\textwidth}{!}{
    \begin{tabular}{cccccccccc}
    \hline
    Backbone & Ratio & 0.025\% & 0.05\% & 0.075\% & 0.1\% & 1\% & 2\% & 4\% & 8\%  \\ \hline
    \multicolumn{1}{c|}{\multirow{4}{*}{MF}} & Retrain & 15 & 17 & 17 & 17 & 16 & 13 & 16 & 12  \\
    \multicolumn{1}{c|}{} & RecEraser & 65 & 69 & 79 & 76 & 65 & 65 & 67 & 72  \\
    \multicolumn{1}{c|}{} & SCIF & 0.014 & 0.004 & 0.005 & 0.006 & 0.006 & 0.013 & 0.008 & 0.012  \\ 
    \multicolumn{1}{c|}{} & IFRU & 0.011 & 0.008 & 0.008 & 0.004 & 0.011 & 0.010 & 0.011 & 0.016  \\ \hline
    
    \multicolumn{1}{c|}{\multirow{4}{*}{LightGCN}} & Retrain & 121 & 168 & 148 & 165 & 173 & 180 & 161 & 175  \\
    \multicolumn{1}{c|}{} & RecEraser & 315 & 368 & 267 & 362 & 318 & 486 & 616 & 365  \\
    \multicolumn{1}{c|}{} & SCIF & 0.1 & 0.2 & 0.4 & 0.3 & 0.4 & 0.5 & 0.5 & 0.6  \\ 
    \multicolumn{1}{c|}{} & IFRU & 0.2 & 0.2 & 0.3 & 0.3 & 0.4 & 0.4 & 0.5 & 0.6\\ \hline
    \end{tabular}
    }
\end{table*} 

\subsubsection{Efficiency Comparison} \label{sec:time-exp}

\begin{figure}[t]
    \centering
    \setlength{\belowcaptionskip}{5mm}
    \subfigure[\textbf{ $AUC_{0}$ --- MF}]{\label{fig:a}\includegraphics[width=0.45\textwidth]{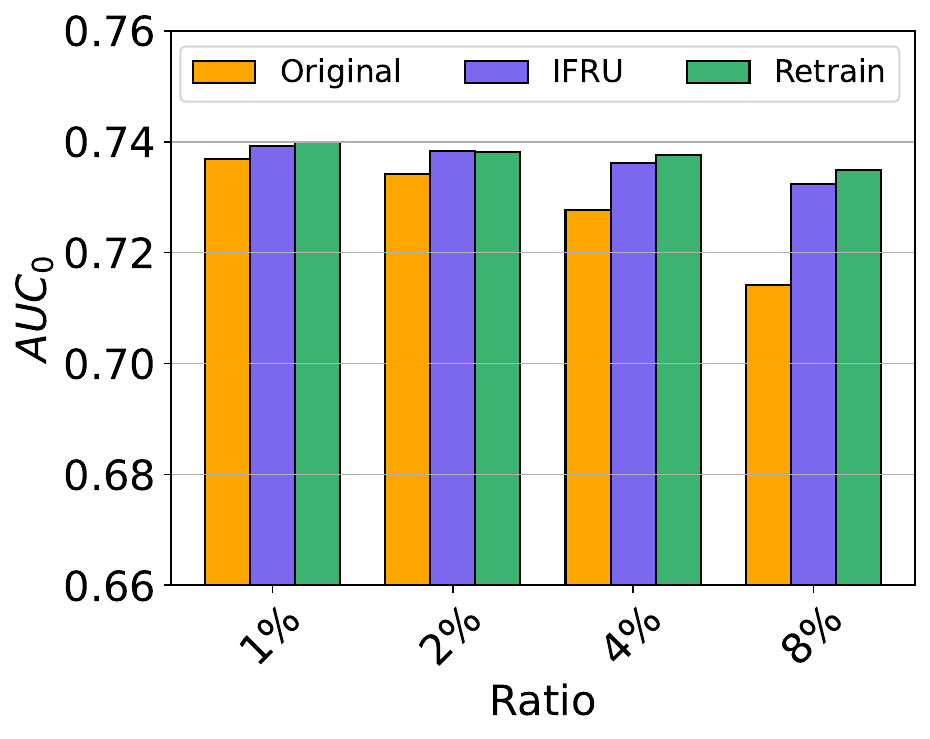}}
    \subfigure[ \textbf{$AUC_{1}$ --- MF}]
    {\label{fig:b} \includegraphics[width=0.45\textwidth]{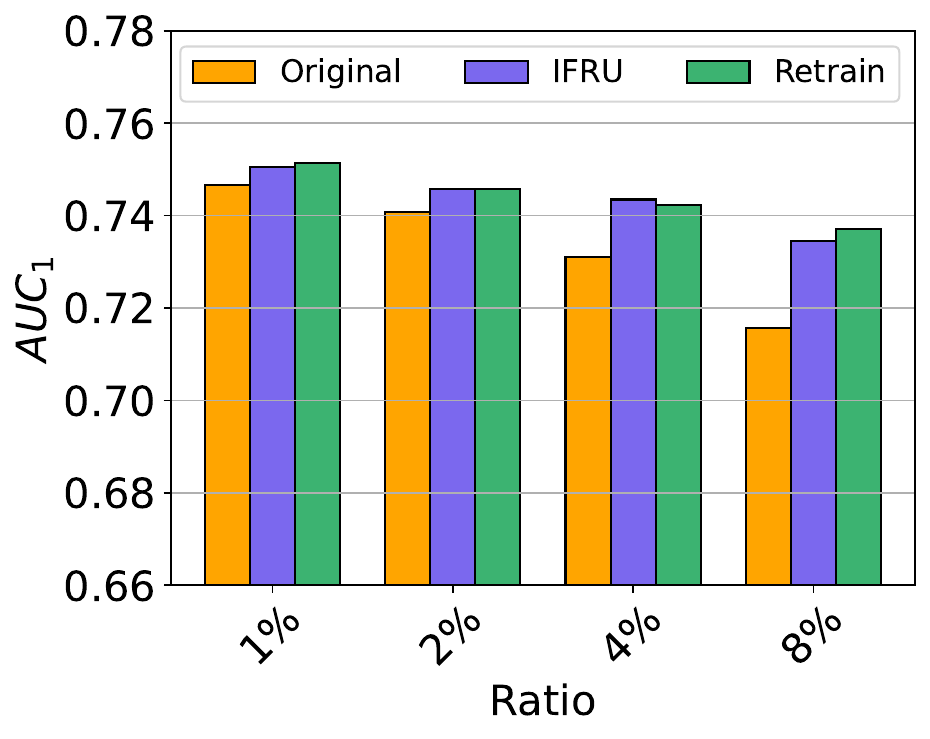}}
    \quad \quad \quad
    \subfigure[\textbf{ $AUC_{0}$ --- LightGCN}]{\label{fig:d}\includegraphics[width=0.45\textwidth]{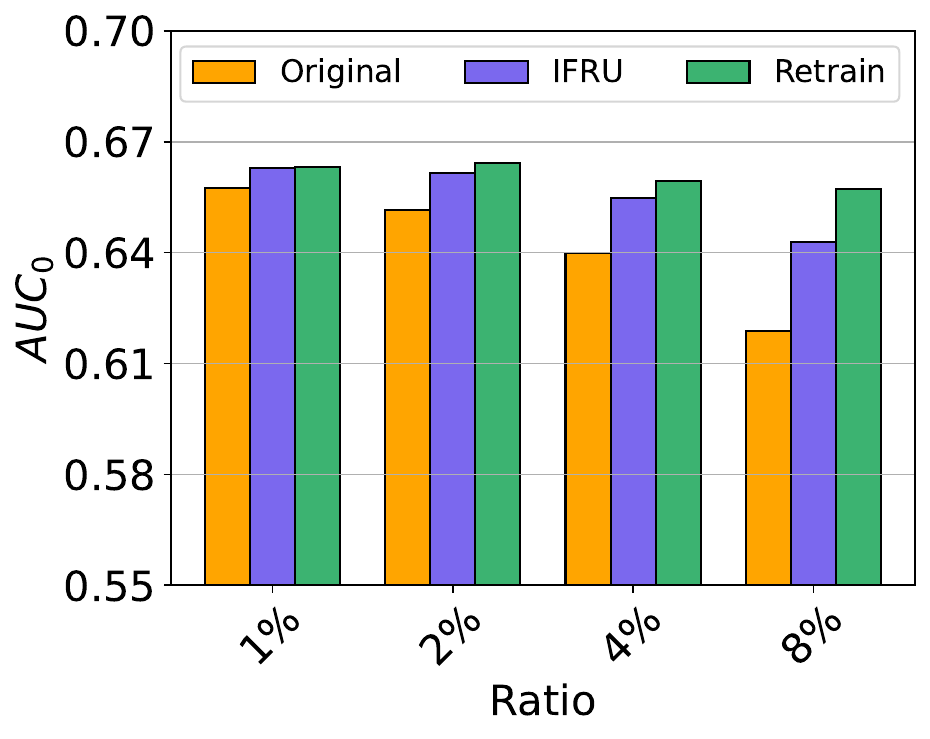}}
    \subfigure[ \textbf{$AUC_{1}$ --- LightGCN}]
    {\label{fig:e} \includegraphics[width=0.45\textwidth]{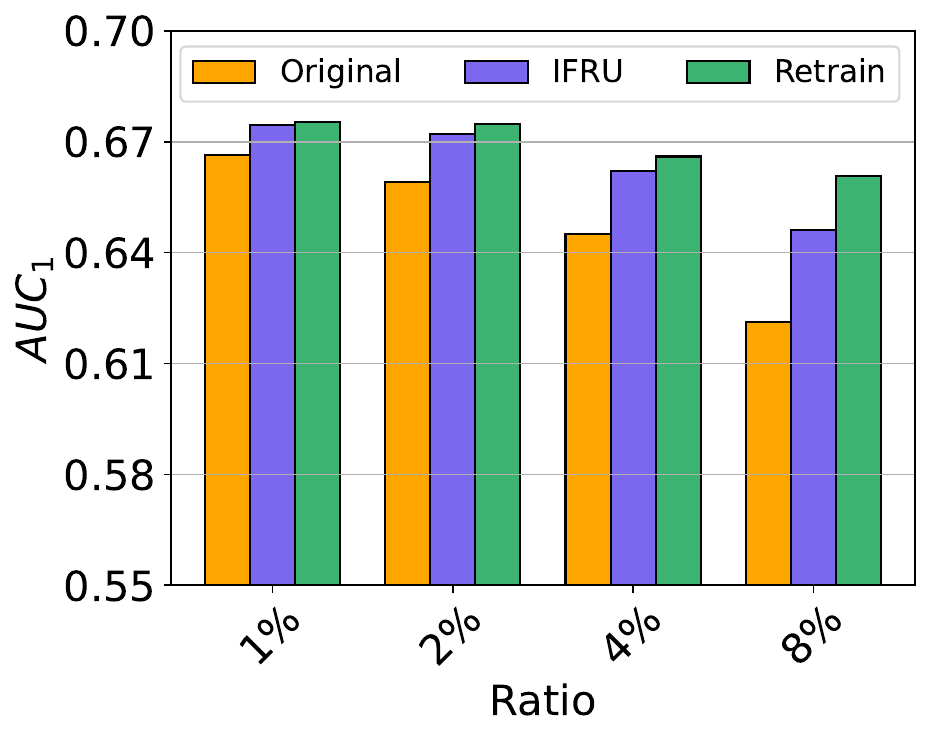}}
    \caption{Performance of Original, Retrain, and IFRU when erasing unusable data of different sizes on the metrics $AUC_{0}$ and $AUC_{1}$ \wrt BookCrossing. `Ratio’ denotes $|\mathcal{D}_{r}|/|\mathcal{D}|$.}
    \label{fig:attack ratios}
    \Description{..}
\end{figure} 

We next compare the running time of IFRU, SCIF, Retrain, and RecEraser to study their efficiency. We omit SISA because it is very similar to RecEraser. To verify the superiority of IFRU in achieving unlearning efficiency, we experiment with erasing unusable data of different sizes. Specially, we control the size according to the ratio of $\mathcal{D}_{r}$ to $\mathcal{D}$ and change the ratio in a wide range of \{8\%, 4\%, 2\%, 1\%, 0.1\%, 0.075\%, 0.05\%, 0.025\%\}. We summarize the results of BookCrossing in Table~\ref{tab:exp_overall_efficiency} and omit the results of Amazon since the trends of results are similar.

According to the table, IFRU and SCIF demonstrate significantly more efficient unlearning compared to Retrain and RecEraser, which can be attributed to their capability to directly modify model parameters without necessitating retraining. For example, IFRU just takes a maximum running time of 0.016s (0.6s) while Retrain and RecEraser take at least hundreds of times longer when using MF (LighGCN) as the backbone model. Notably, SCIF sometimes appears more efficient than our method, this is because it does not account for spillover influence. Regarding RecEraser, which is designed to pursue efficient unlearning,  we find it indeed cannot improve unlearning efficiency compared to Retrain, and even becomes worse\footnote{The conclusion differs from the original RecEraser paper as it dealt with sequentially arriving unlearning requests, whereas our work focuses on a more realistic batch setting where unusable data is randomly distributed and needs to be removed simultaneously.}. This is because RecEraser accelerates unlearning only if the unusable data is distributed in a few data shards split by RecEraser, which is not satisfied in our setting. For example, even if the ratio is only $0.025\%$, the randomly sampled unusable data is distributed in all data shards. Different from RecEraser, our IFRU could achieve efficient unlearning without limiting the size and distribution of unusable data, making it more suitable for practical scenarios where the unusable data is diverse and huge.

\begin{figure}[t]
    \centering
    \subfigure[\textbf{Amazon}]{\label{fig:g}\includegraphics[width=0.45\textwidth]{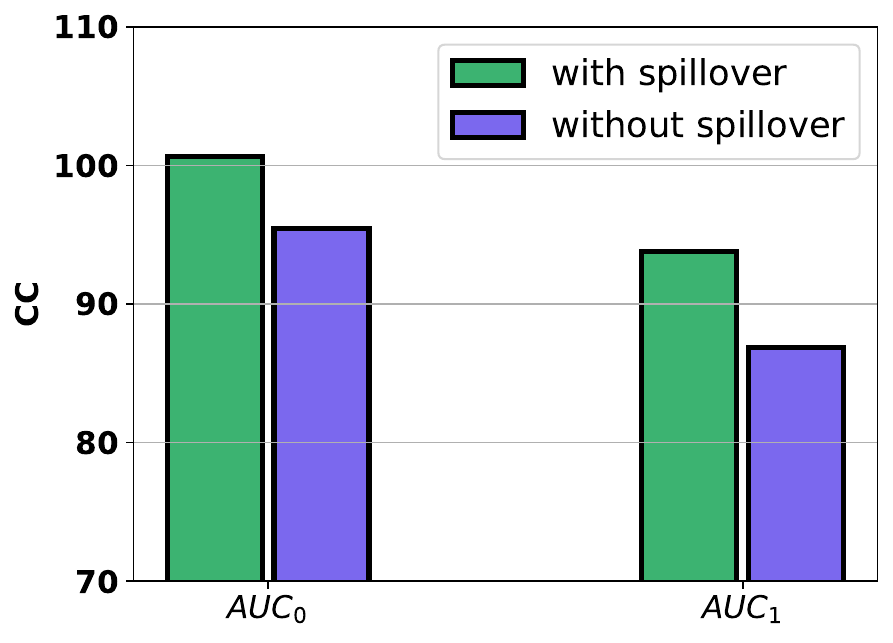}}
    \subfigure[\textbf{BookCrossing}]{\label{fig:h} \includegraphics[width=0.45\textwidth]{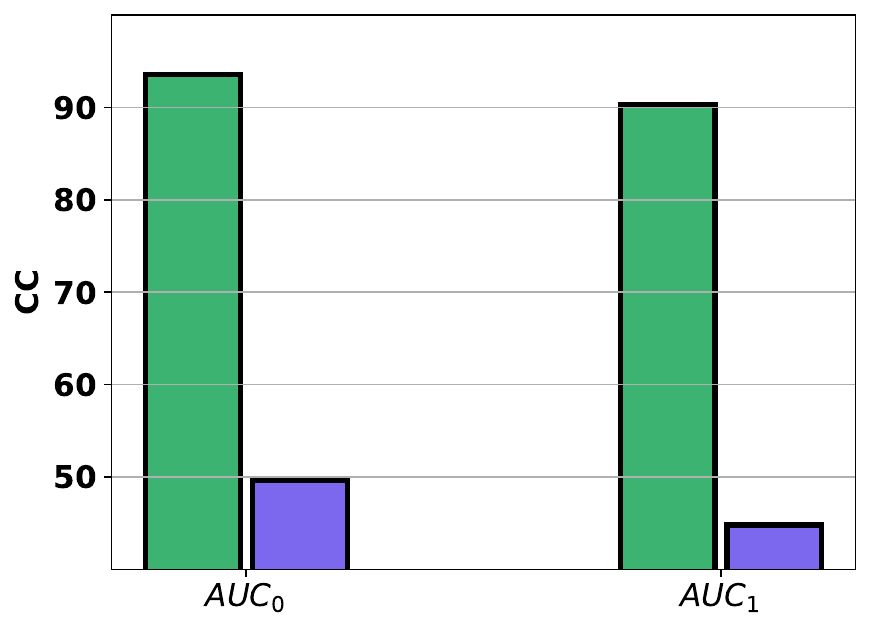}}
       \caption{Comparisons between IFRU with and without spillover influence on the backbone model LightGCN \wrt $AUC_0$ and $AUC_1$, with values closer to 100\% indicating complete unlearning. }
    \label{fig:spillover-final}
    \Description{..}
\end{figure}

\subsection{In-depth Studies (RQ2)}

We next conduct some in-depth analyses to understand how the unlearning settings (the ratio of unusable data) and our method designs (spillover influence, pruning, and HVP) impact the effectiveness and efficiency of the proposed IFRU.

\subsubsection{Effect of the Ratio of Unusable Data}
In the preceding section, we have studied the running time of IFRU when the ratio of unusable data varies. We next study the impact of the ratio of unusable data on the unlearning completeness. If taking a small ratio of unusable data, the performance gap caused by erasing the data would be much smaller, resulting in difficulties in analyzing the unlearning completeness. Therefore, we only keep the ratios of unusable data --- \{1\%, 2\%, 4\%, 8\%\} and ignore small ratios. Figure~\ref{fig:attack ratios} summarizes the results of Original, Retrain, and IFRU under these ratios \wrt the metrics $AUC_{0}$ and ${AUC_{1}}$. 

From the figure, we draw the following observations. 1) Increasing the ratio of unusable data (\ie attack data in this work) reduces the performances of Original and Retrain. Original uses all training data to train, so increasing the ratio of attack data would make it more susceptible to the attack. Although Retrain uses the remaining clean data for model training, the size of clean data (\ie $|\mathcal{D}_{-r}|$) shrinks as the ratio of unusable data increases and hence the performance of Retain also decreases. 2) The performance of IFRU is close to that of Retrain. We take the completeness coefficient (\cf Section~\ref{sec: acc-comp}) to quantify the completeness of unlearning again. The minimal coefficient of IFRU is over $85\%$ for all metrics (including $AUC_{2}$) under all ratios on the MF backbone. On the LightGCN backbone, the minimal coefficient of IFRU is close to $80\%$ for all metrics (including $AUC_{2}$) under the ratio from $1\%$ to $4\%$. These results suggest that IFRU can achieve relatively complete unlearning, even if erasing a large ratio of unusable data.

\subsubsection{Effect of Spillover Influence}

According to the analysis in Section~\ref{Method}, it is essential to estimate spillover influence to achieve complete unlearning. Here we provide empirical evidence by comparing IFRU with a variant of IFRU that discards spillover influence. We conduct experiments on both Amazon and BookCrossing with LightGCN as the backbone model and take the same setting to Section~\ref{sec: acc-comp}. We still take the  completeness coefficient defined in Section~\ref{sec: acc-comp} to quantify how complete the unlearning is. Figure~\ref{fig:spillover-final} shows the coefficient \wrt $AUC_{0}$ and $AUC_1$. We can observe that on both metrics, the coefficient obtained by IFRU without spillover influence is far less than the one considering spillover influence, especially for BookCrossing, indicating more incomplete unlearning. 
This verifies the importance of estimating spillover influence to achieve complete unlearning.

\subsubsection{Effect of Pruning}\label{sec:pruning-exp}

\begin{figure}[t]
    \centering
    \subfigure[\textbf{$AUC_0$---MF}]{\includegraphics[width=0.45\textwidth]{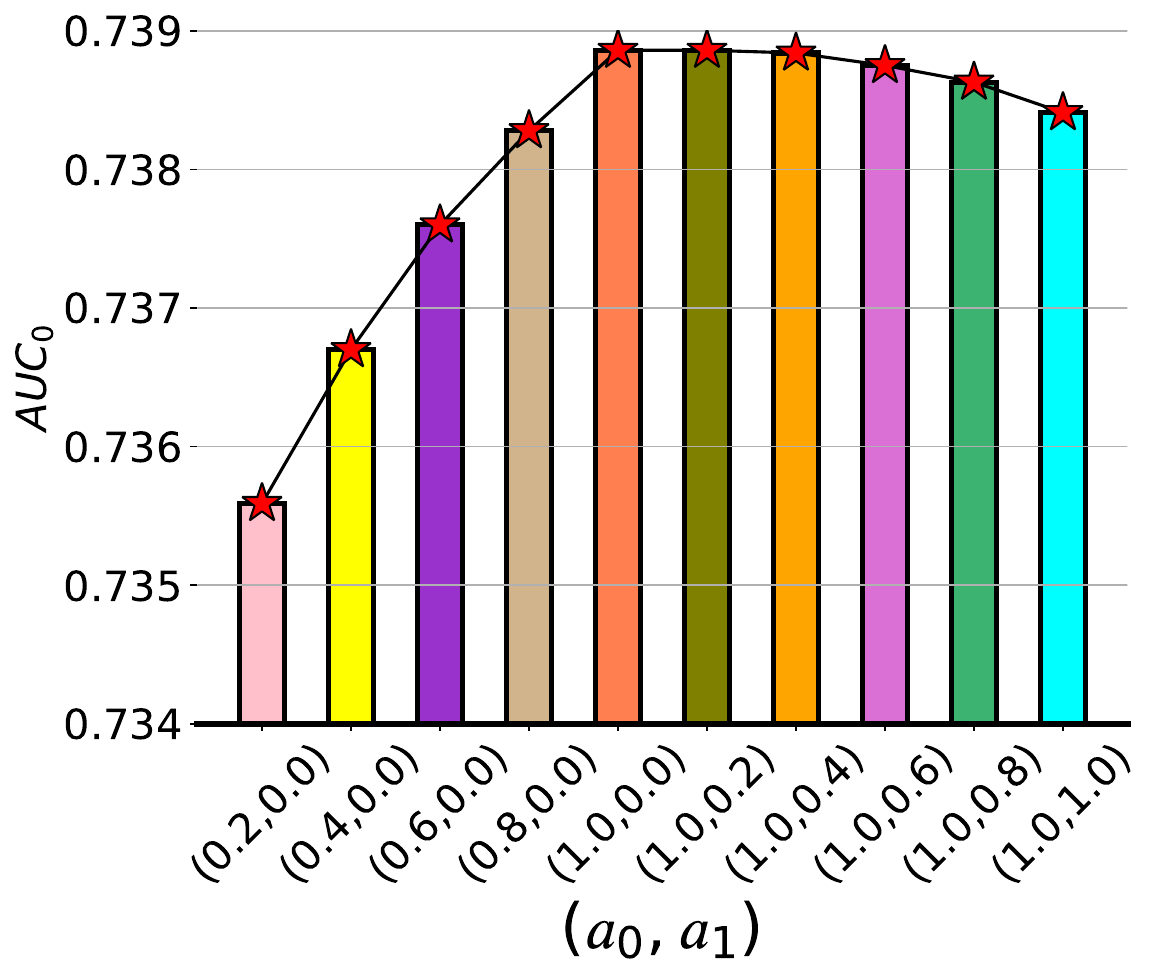}}
    \subfigure[\textbf{Running time---MF}]{\label{fig:i}\includegraphics[width=0.45\textwidth]{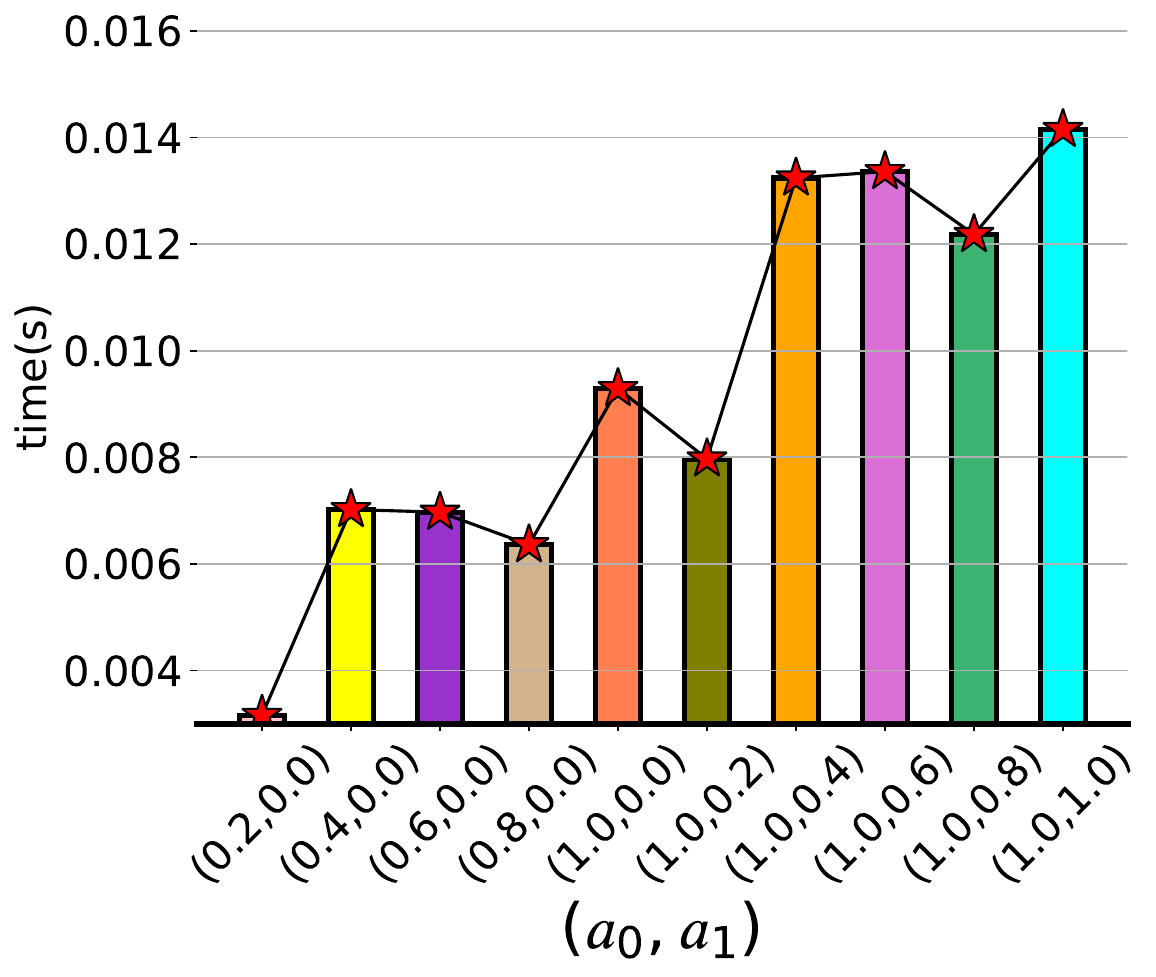}}
    \quad \quad \quad \quad
    \subfigure[ \textbf{ $AUC_0$---LightGCN}]{ \includegraphics[width=0.45\textwidth]{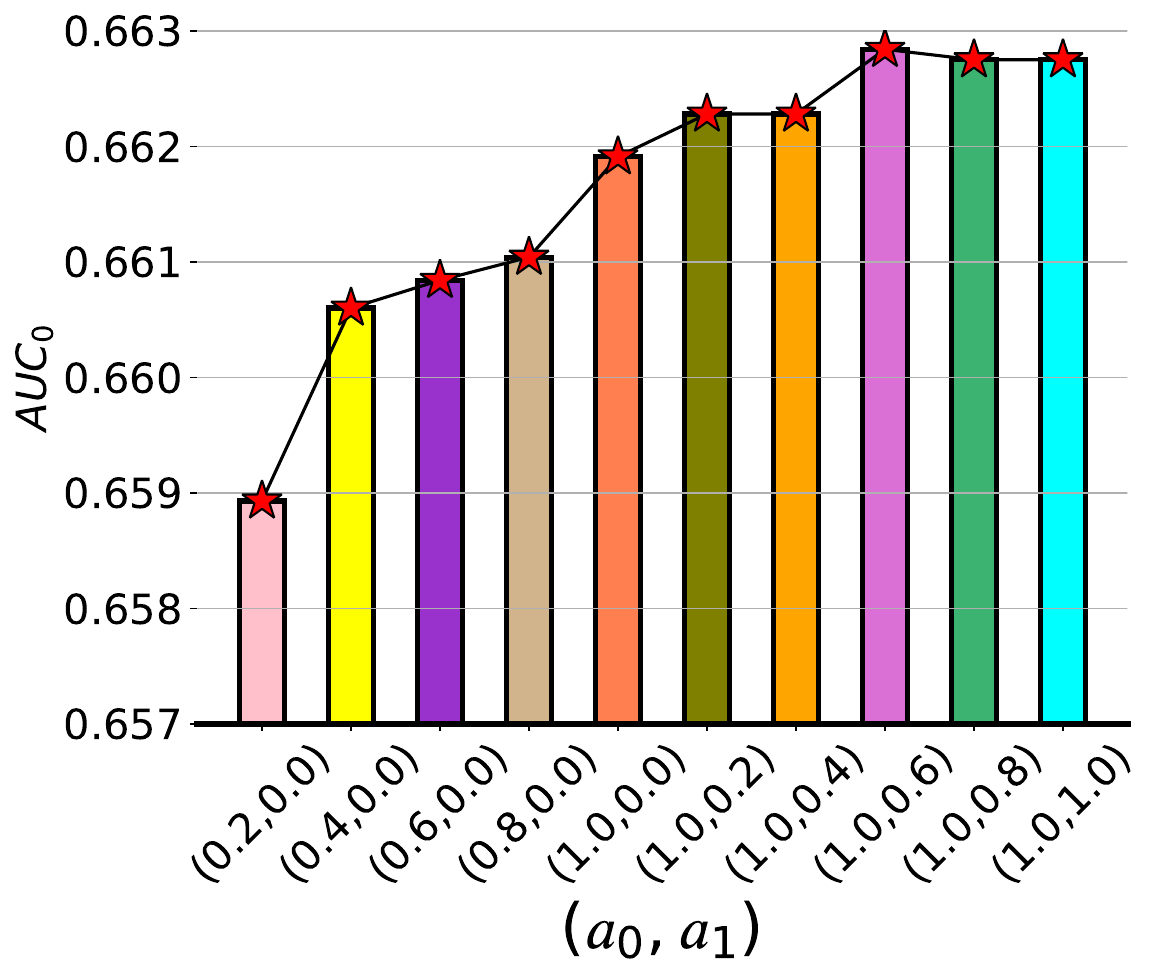}}
    \subfigure[ \textbf{Running time ---LightGCN}]{\label{fig:j} \includegraphics[width=0.45\textwidth]{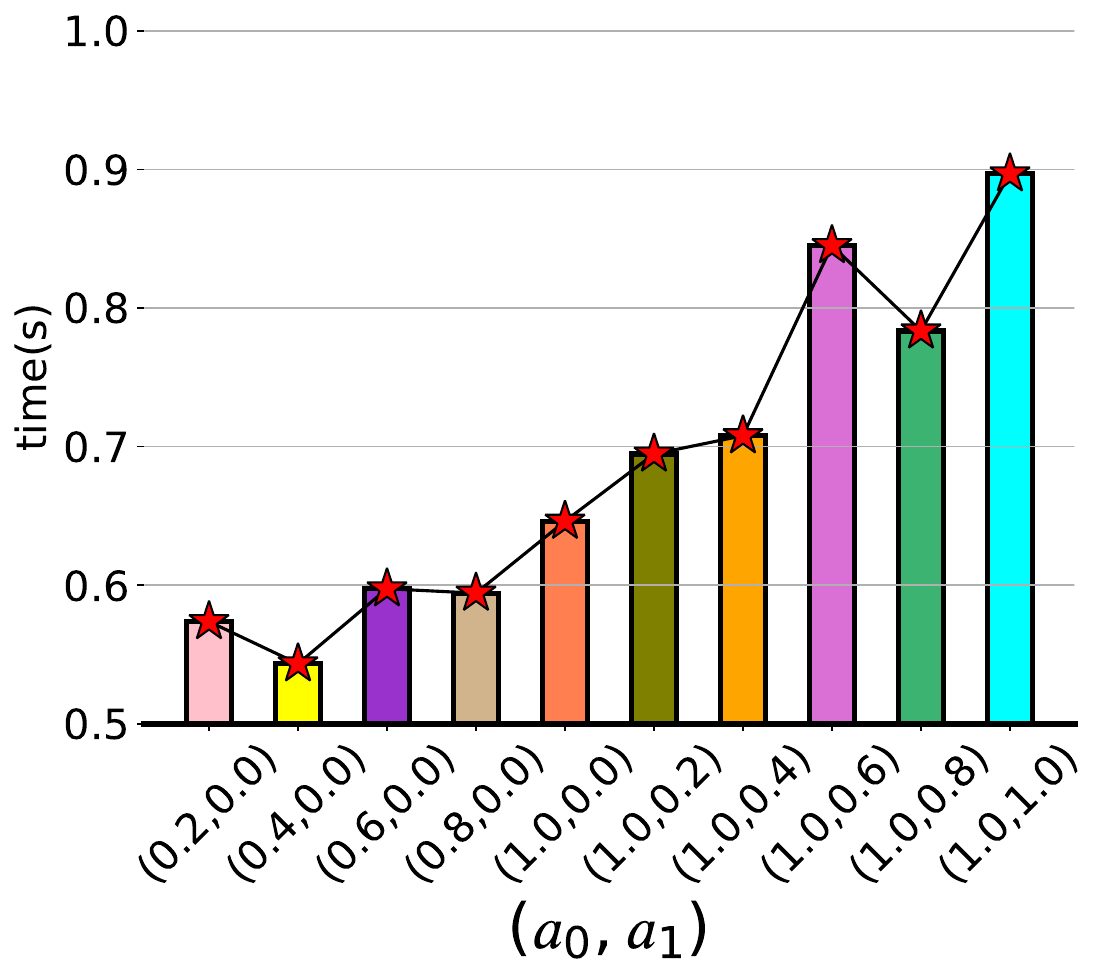}}
    \caption{Recommendation accuracy, measured by $AUC_{0}$, and running time of IFRU \wrt the pruning ratio $(a_{0}, a_{1})$ on BookCrossing. Results for other accuracy metrics show similar trends and are thus omitted. The time cost evaluation was performed on a machine with one NVIDIA RTX 3060 GPU.}
    \label{fig:pruning-studies}
    \Description{..}
\end{figure}

To illustrate the effectiveness of the proposed pruning algorithm, we study the impact of the pruning ratios on unlearning efficiency and completeness. The experiments are conducted on BookCrossing using MF and LightGCN as the backbone models. We take the default experimental settings except for changing the ratio $(a_0,a_1)$ of Algorithm~\ref{alg:pruning}. Specially, we vary both $a_{0}$ and $a_{1}$ in the range of \{0.0, 0.2, 0.4, 0.6, 0.8, 1.0\}, and keep $a_{0}=1$ if  $a_{1}$ is greater than $0$. That means we start pruning the $0$-order neighbors only if prune all $1$-order neighbors in Algorithm~\ref{alg:pruning}. After ignoring (0.0, 0.0) which means pruning all model parameters, $(1.0,1.0)$ means minimal pruning, and $(0.2,0)$ means maximum pruning. We summarize the results of $AUC_{0}$ and running time in Figure~\ref{fig:pruning-studies}. We omit the results of $AUC_{1}$ and $AUC_{2}$ to save space, considering that they show similar trends to $AUC_{0}$.

From the results in Figure~\ref{fig:pruning-studies}, we have the following observations. First, as pruning more model parameters (from the right to the left in the figure), the running time of IFRU shows decreasing trends. The decreasing trend on MF is relatively unstable, because the running time on MF is very small, which is sensitive to running environments. Second, we find the result on $AUC_{0}$ only slightly varies before we start pruning the $0$-order neighbors (\ie before going to the left of $(1.0,0.0)$ in the figure). Note that the ratio of $(1.0,0.0)$ corresponds to pruning at least $5/6$ of all model parameters. These results indicate that our pruning could accelerate unlearning with small sacrifices of unlearning completeness even if pruning a large ratio of model parameters, verifying the effectiveness of our method in pruning less important model parameters. 

\subsubsection{Effect of HVP}

Efficient calculation of the influence $I(\hat{\theta},\mathcal{D}_{r})$ depends on the use of the Hessian Vector Product (HVP) in Equation~\eqref{eq:hvp}, as discussed in Section~\ref{sec:calculating}. In this subsection, we investigate the impact of HVP on the efficacy and efficiency of IFRU. To achieve this goal, we propose a variant of IFRU called IFRU without HVP (IFRU w/o HVP). This variant computes the Hessian matrix in Equation~\eqref{eq:optimize-influence} explicitly in advance to calculate the influence. We compare the performance of IFRU w/o HVP with that of the original IFRU with HVP. For the comparison, we use the default experimental settings and the same environment as Section~\ref{sec:pruning-exp}, except for the pruning ratios $(a_0,a_1)$ of Algorithm~\ref{alg:pruning} for LightGCN. We set $(a_0,a_1)$ to (1.0, 0.2), as the default setting of (1.0, 0.6) leads to an out-of-memory issue for IFRU w/o HVP during the loading of the Hessian matrix for LightGCN, even if represented as a sparse matrix. We present the comparison results for the BookCrossing dataset in Table~\ref{tab:hvp}, which includes both the recommendation accuracy and the running time of the unlearning process.

The results in Table~\ref{tab:hvp} show that applying HVP to IFRU significantly accelerates the unlearning process without any negative impact on recommendation accuracy. On the other hand, IFRU w/o HVP is slower than the baselines, primarily due to its much higher I/O costs. As the Hessian matrix is very large, we can only load it onto the GPU in a mini-batch manner and need to repeatedly load it in each iteration while calculating the influence according to Equation~\eqref{eq:optimize-influence}, which results in a significant time cost. 

\begin{table}[t]
    \centering
    \caption{Recommendation accuracy (measured by $AUC_{0}$ and $AUC_{1}$) and running time of IFRU with HVP (IFRU w/ HVP) and IFRU without HVP (IFRU w/o HVP) on the BookCrossing dataset. The time cost evaluation was performed on a machine with one NVIDIA RTX 3060 GPU for all methods.}
    \label{tab:hvp}
    \resizebox{0.77\textwidth}{!}{%
    \begin{tabular}{c|ccc|ccc}
    \hline
    Backbone & \multicolumn{3}{c|}{MF} & \multicolumn{3}{c}{LightGCN} \\ \hline
    Methods & $AUC_0$ & $AUC_1$ & Time {[}s{]} & $AUC_0$ & $AUC_1$ & Time {[}s{]} \\ \hline
    IFRU w/ HVP & 0.7384 & 0.7458 & 0.014 & 0.6623 & 0.6731 & 0.7 \\
    IFRU w/o HVP & 0.7384 & 0.7458 & 51 & 0.6623 & 0.6731 & 520 \\ \hline
    \end{tabular}
    }
\end{table}
\section{Related Work}

This work is closely related to two topics: machine unlearning and influence function in recommendation. In the following, we discuss the related works that have been done on these topics.

\subsection{Machine Unlearning}
\textit{Machine unlearning.} In machine learning, trained models undoubtedly memorize the information of training data~\cite{machineUnlearn,survey2}. In some cases, we need to remove the influence of some training data on the model parameters due to various reasons~\cite{nguyen2022survey,survey2}, \eg privacy~\cite{cao2015towards} and security~\cite{marchant2022hard}. The task is called machine unlearning, and there appear two types of methods: data-driven and non-data-driven approaches~\cite{nguyen2022survey}. The latter is less related to us so we do not discuss here.

Regarding data-driven methods, there are three lines of work. The first line of work is based on retraining~\cite{machineUnlearn,chen2022graph}, which achieves erasing data by retraining but designs new model architectures or training paradigms. Due to the modifications to the model or training, these methods may lead to reduced accuracy, thus hampering the effectiveness of harmless unlearning. The second line of work achieves unlearning by data augmentation~\cite{shan2020fawkes,tarun2021fast}. For example, \cite{tarun2021fast} generates error-maximizing noise to manipulate the model weights to unlearn the targeted class of data. This method avoids excessive retention of redundant intermediate data. However, when complex models are subject to unlearning requests, it can be challenging to offset the impact of such unlearning data completely. The third line of work achieves unlearning via influence function~\cite{PUMA,inf-unlearn,GIF}. \cite{PUMA} models the influence of each training data point on the model concerning various performance criteria (beyond training loss) and then removes the negative impact. \cite{inf-unlearn} extends the influence function to estimate the influence of features for achieving feature unlearning. This unlearning approach directly manipulates the existing model without retraining, making it more efficient, but may involve approximation errors if its assumptions are not well satisfied. Our unlearning method falls within the third category of research, with a specific emphasis on recommendation unlearning. In this context, we extend the influence function to encompass the modeling of spillover influence.

In addition to researching how to achieve unlearning, there are also some works that specifically focus on the benchmarks~\cite{schelter8364amnesia}, applications~\cite{unlean-app}, verification~\cite{huang2021ema,thudi2022unrolling}, and security~\cite{thudi2022necessity} of machine unlearning. However, research on these aspects is still in its early stages and is considered open problems~\cite{survey2,machineUnlearn}. Although these aspects are undoubtedly important, our work is less related to these directions. As a result, we do not discuss these aspects in detail in this paper, but we acknowledge their significance and leave them for future research.

\vspace{+5pt}
\textit{Recommendation unlearning}. Recently, recommendation unlearning receives certain attention~\cite{yuan2022federated,Recunlearn,LASER,liu2022forgetting,xu2023netflix}. However, these methods achieve unlearning by retraining but consider different strategies to speed up the retraining. By splitting the training data/model into shards/sub-models, \cite{Recunlearn,LASER} just retrain a few sub-models from scratch to accelerate unlearning, assuming the unusable data only affects a few sub-models. However, the assumption is not practical. \cite{liu2022forgetting} accelerates unlearning by starting retraining in a warm-start way and using quasi-Newton optimization for retraining. However, it is a method similar to fine-tuning, which is not approved for unlearning. The approach proposed in~\cite{xu2023netflix} is also a fine-tuning method, but it is specifically designed for Alternating Least Squares (ALS)~\cite{takacs2011applications} and bi-linear models.  \cite{yuan2022federated} is designed for the federated recommendation. To accelerate unlearning, it tries to reduce the size of retraining data by roughly approximating retraining gradients with the initial training gradients. The approximation undoubtedly has side effects on the performance. Different from these methods, we achieve unlearning without retraining via the influence function. Our method is efficient naturally and the unlearning is also relatively complete. Moreover, our method does not involve modifications to model/learning architectures, having no side effects on model performance. Apart from these methods, we have noticed a recent work~\cite{li2023selective} that also uses the influence function to achieve recommendation unlearning. However, the work~\cite{li2023selective} does not take into account the information of high-order neighbors and the spillover effects that arise from changes in the computational graph due to removing the unusable data. We emphasize the need to consider these effects in order to achieve more complete unlearning.

\subsection{Influence Function in Recommendation}
\textit{Influence function} is a classic concept originating from robust statistics~\cite{hampel1974influence}, which is used to estimate the influence of data on learning models. It has been widely used for various tasks in various areas~\cite{zhang2021sample,feldman2020neural,IF}. After \cite{IF} utilizes it to achieve great success in explaining the prediction of black-box models, it has been widely used for various tasks in various areas~\cite{zhang2021sample,feldman2020neural}. Moreover, a number of studies~\cite{bae2022if,fragile} have conducted detailed analyses on the characteristics of the influence function in deep learning, including the estimation error. This has led to a contentious debate surrounding its efficacy in deep learning~\cite{fragile}. However, despite the ongoing controversy, a significant body of recent research continues to utilize the influence function and provides empirical evidence of its effectiveness~\cite{bae2022if,GIF,martinez2023approximating, Nguyen-DucTTHND23}.

In recommendation, the influence function has also been utilized for dealing with various tasks in recent research. First, some efforts~\cite{zhang2020practical,wu2021triple,fang2020influence,yi2014robust} apply the influence function to achieve more efficient and effective recommendation attacks. For example, ~\cite{wu2021triple} utilizes the influence function to quickly estimate the quality of generated fake users, helping find more valid fake users; and~\cite{fang2020influence} utilizes the influence function to identify more influential users that are more important for the prediction of a target item, and then focus on attacking these users to achieve better results. Second, the influence function has also been used for debiasing~\cite{yu2020influence}. Specifically, ~\cite{yu2020influence} utilizes the influence function to quantify how training points are essential for the model performance on unbiased data and further takes the estimated importance scores to weight training data for debiasing. Besides, the influence function has also been utilized to explain the recommendation results in \cite{cheng2019incorporating}, which first measures the influences of training interactions on predictions and then provides neighbor-style explanations based on the most influential training interactions. Different from these researches, we take influence function to achieve recommendation unlearning. Additionally, we extend the influence function to model the spillover influence, which makes it more appropriate for recommendation systems compared to solely applying the influence function. 
\section{Conclusion}

In this work, we study how to achieve complete, efficient unlearning without side effects on recommendation performance via the influence function. Towards this goal, we propose a novel and general unlearning method, which extends the influence function to estimate the spillover influence of unusable data to achieve more complete unlearning, and prunes less important parameters to further improve the unlearning efficiency. We conduct extensive experiments on two real-world datasets, providing insightful analysis for the effectiveness and great efficiency of our proposal in erasing interaction data. In the future, we will extend our method to deal with other types of data, \eg sensitive features instead of interactions. Furthermore, we plan to apply our method to a diverse range of recommender models, such as factorization machines~\cite{rendleFM, DIL} and sequential models~\cite{seq-rec}, which can incorporate various side information. In particular, recommendation systems based on large language models have received significant attention recently~\cite{wu2023survey,bao2023tallrec,lin2023can}, and the massive parameters of such models pose new efficiency and efficacy challenges for existing unlearning approaches~\cite{hu2024exact}, including our proposed method. Hence, we will  explore how to apply our method to these large models as well.

\begin{acks}
This work is supported by the National Natural Science Foundation of China (62272437).
\end{acks}

\bibliographystyle{ACM-Reference-Format}
% \balance
\bibliography{8_reference}


\begin{thebibliography}{60}

%%% the Web address.

\ifx \showCODEN    \undefined \def \showCODEN     #1{\unskip}     \fi
\ifx \showDOI      \undefined \def \showDOI       #1{#1}\fi
\ifx \showISBNx    \undefined \def \showISBNx     #1{\unskip}     \fi
\ifx \showISBNxiii \undefined \def \showISBNxiii  #1{\unskip}     \fi
\ifx \showISSN     \undefined \def \showISSN      #1{\unskip}     \fi
\ifx \showLCCN     \undefined \def \showLCCN      #1{\unskip}     \fi
\ifx \shownote     \undefined \def \shownote      #1{#1}          \fi
\ifx \showarticletitle \undefined \def \showarticletitle #1{#1}   \fi
\ifx \showURL      \undefined \def \showURL       {\relax}        \fi
% invisible to TeX
\providecommand\bibfield[2]{#2}
\providecommand\bibinfo[2]{#2}
\providecommand\natexlab[1]{#1}
\providecommand\showeprint[2][]{arXiv:#2}

\bibitem[Bae et~al\mbox{.}(2022)]%
        {bae2022if}
\bibfield{author}{\bibinfo{person}{Juhan Bae}, \bibinfo{person}{Nathan Ng}, \bibinfo{person}{Alston Lo}, \bibinfo{person}{Marzyeh Ghassemi}, {and} \bibinfo{person}{Roger~B Grosse}.} \bibinfo{year}{2022}\natexlab{}.
\newblock \showarticletitle{If Influence Functions are the Answer, Then What is the Question?}
\newblock \bibinfo{journal}{\emph{Advances in Neural Information Processing Systems}}  \bibinfo{volume}{35} (\bibinfo{year}{2022}), \bibinfo{pages}{17953--17967}.
\newblock


\bibitem[Bao et~al\mbox{.}(2023)]%
        {bao2023tallrec}
\bibfield{author}{\bibinfo{person}{Keqin Bao}, \bibinfo{person}{Jizhi Zhang}, \bibinfo{person}{Yang Zhang}, \bibinfo{person}{Wenjie Wang}, \bibinfo{person}{Fuli Feng}, {and} \bibinfo{person}{Xiangnan He}.} \bibinfo{year}{2023}\natexlab{}.
\newblock \showarticletitle{Tallrec: An effective and efficient tuning framework to align large language model with recommendation}. In \bibinfo{booktitle}{\emph{Proceedings of the 17th ACM Conference on Recommender Systems}}. \bibinfo{pages}{1007--1014}.
\newblock


\bibitem[Basu et~al\mbox{.}(2021)]%
        {fragile}
\bibfield{author}{\bibinfo{person}{Samyadeep Basu}, \bibinfo{person}{Phillip Pope}, {and} \bibinfo{person}{Soheil Feizi}.} \bibinfo{year}{2021}\natexlab{}.
\newblock \showarticletitle{Influence Functions in Deep Learning Are Fragile}. In \bibinfo{booktitle}{\emph{9th International Conference on Learning Representations}}.
\newblock


\bibitem[Bourtoule et~al\mbox{.}(2021)]%
        {machineUnlearn}
\bibfield{author}{\bibinfo{person}{Lucas Bourtoule}, \bibinfo{person}{Varun Chandrasekaran}, \bibinfo{person}{Christopher~A Choquette-Choo}, \bibinfo{person}{Hengrui Jia}, \bibinfo{person}{Adelin Travers}, \bibinfo{person}{Baiwu Zhang}, \bibinfo{person}{David Lie}, {and} \bibinfo{person}{Nicolas Papernot}.} \bibinfo{year}{2021}\natexlab{}.
\newblock \showarticletitle{Machine unlearning}. In \bibinfo{booktitle}{\emph{2021 IEEE Symposium on Security and Privacy (SP)}}. IEEE, \bibinfo{pages}{141--159}.
\newblock


\bibitem[Cao and Yang(2015)]%
        {cao2015towards}
\bibfield{author}{\bibinfo{person}{Yinzhi Cao} {and} \bibinfo{person}{Junfeng Yang}.} \bibinfo{year}{2015}\natexlab{}.
\newblock \showarticletitle{Towards making systems forget with machine unlearning}. In \bibinfo{booktitle}{\emph{2015 IEEE Symposium on Security and Privacy}}. IEEE, \bibinfo{pages}{463--480}.
\newblock


\bibitem[Chen et~al\mbox{.}(2022a)]%
        {Recunlearn}
\bibfield{author}{\bibinfo{person}{Chong Chen}, \bibinfo{person}{Fei Sun}, \bibinfo{person}{Min Zhang}, {and} \bibinfo{person}{Bolin Ding}.} \bibinfo{year}{2022}\natexlab{a}.
\newblock \showarticletitle{Recommendation unlearning}. In \bibinfo{booktitle}{\emph{Proceedings of the ACM Web Conference 2022}}. \bibinfo{pages}{2768--2777}.
\newblock


\bibitem[Chen et~al\mbox{.}(2022b)]%
        {chen2022graph}
\bibfield{author}{\bibinfo{person}{Min Chen}, \bibinfo{person}{Zhikun Zhang}, \bibinfo{person}{Tianhao Wang}, \bibinfo{person}{Michael Backes}, \bibinfo{person}{Mathias Humbert}, {and} \bibinfo{person}{Yang Zhang}.} \bibinfo{year}{2022}\natexlab{b}.
\newblock \showarticletitle{Graph unlearning}. In \bibinfo{booktitle}{\emph{Proceedings of the 2022 ACM SIGSAC Conference on Computer and Communications Security}}. \bibinfo{pages}{499--513}.
\newblock


\bibitem[Cheng et~al\mbox{.}(2019)]%
        {cheng2019incorporating}
\bibfield{author}{\bibinfo{person}{Weiyu Cheng}, \bibinfo{person}{Yanyan Shen}, \bibinfo{person}{Linpeng Huang}, {and} \bibinfo{person}{Yanmin Zhu}.} \bibinfo{year}{2019}\natexlab{}.
\newblock \showarticletitle{Incorporating interpretability into latent factor models via fast influence analysis}. In \bibinfo{booktitle}{\emph{Proceedings of the 25th ACM SIGKDD International Conference on Knowledge Discovery \& Data Mining}}. \bibinfo{pages}{885--893}.
\newblock


\bibitem[Cook and Weisberg(1982)]%
        {cook1982residuals}
\bibfield{author}{\bibinfo{person}{R~Dennis Cook} {and} \bibinfo{person}{Sanford Weisberg}.} \bibinfo{year}{1982}\natexlab{}.
\newblock \bibinfo{booktitle}{\emph{Residuals and influence in regression}}.
\newblock \bibinfo{publisher}{New York: Chapman and Hall}.
\newblock


\bibitem[Fang et~al\mbox{.}(2020b)]%
        {seq-rec}
\bibfield{author}{\bibinfo{person}{Hui Fang}, \bibinfo{person}{Danning Zhang}, \bibinfo{person}{Yiheng Shu}, {and} \bibinfo{person}{Guibing Guo}.} \bibinfo{year}{2020}\natexlab{b}.
\newblock \showarticletitle{Deep learning for sequential recommendation: Algorithms, influential factors, and evaluations}.
\newblock \bibinfo{journal}{\emph{ACM Transactions on Information Systems (TOIS)}} \bibinfo{volume}{39}, \bibinfo{number}{1} (\bibinfo{year}{2020}), \bibinfo{pages}{1--42}.
\newblock


\bibitem[Fang et~al\mbox{.}(2020a)]%
        {fang2020influence}
\bibfield{author}{\bibinfo{person}{Minghong Fang}, \bibinfo{person}{Neil~Zhenqiang Gong}, {and} \bibinfo{person}{Jia Liu}.} \bibinfo{year}{2020}\natexlab{a}.
\newblock \showarticletitle{Influence function based data poisoning attacks to top-n recommender systems}. In \bibinfo{booktitle}{\emph{Proceedings of The Web Conference 2020}}. \bibinfo{pages}{3019--3025}.
\newblock


\bibitem[Fawcett(2006)]%
        {fawcett2006introduction}
\bibfield{author}{\bibinfo{person}{Tom Fawcett}.} \bibinfo{year}{2006}\natexlab{}.
\newblock \showarticletitle{An introduction to ROC analysis}.
\newblock \bibinfo{journal}{\emph{Pattern recognition letters}} \bibinfo{volume}{27}, \bibinfo{number}{8} (\bibinfo{year}{2006}), \bibinfo{pages}{861--874}.
\newblock


\bibitem[Feldman and Zhang(2020)]%
        {feldman2020neural}
\bibfield{author}{\bibinfo{person}{Vitaly Feldman} {and} \bibinfo{person}{Chiyuan Zhang}.} \bibinfo{year}{2020}\natexlab{}.
\newblock \showarticletitle{What neural networks memorize and why: Discovering the long tail via influence estimation}.
\newblock \bibinfo{journal}{\emph{Advances in Neural Information Processing Systems}}  \bibinfo{volume}{33} (\bibinfo{year}{2020}), \bibinfo{pages}{2881--2891}.
\newblock


\bibitem[Hampel(1974)]%
        {hampel1974influence}
\bibfield{author}{\bibinfo{person}{Frank~R Hampel}.} \bibinfo{year}{1974}\natexlab{}.
\newblock \showarticletitle{The influence curve and its role in robust estimation}.
\newblock \bibinfo{journal}{\emph{Journal of the american statistical association}} \bibinfo{volume}{69}, \bibinfo{number}{346} (\bibinfo{year}{1974}), \bibinfo{pages}{383--393}.
\newblock


\bibitem[He and McAuley(2016)]%
        {amazon-data}
\bibfield{author}{\bibinfo{person}{Ruining He} {and} \bibinfo{person}{Julian McAuley}.} \bibinfo{year}{2016}\natexlab{}.
\newblock \showarticletitle{Ups and downs: Modeling the visual evolution of fashion trends with one-class collaborative filtering}. In \bibinfo{booktitle}{\emph{proceedings of the 25th international conference on world wide web}}. \bibinfo{pages}{507--517}.
\newblock


\bibitem[He et~al\mbox{.}(2020)]%
        {lightgcn}
\bibfield{author}{\bibinfo{person}{Xiangnan He}, \bibinfo{person}{Kuan Deng}, \bibinfo{person}{Xiang Wang}, \bibinfo{person}{Yan Li}, \bibinfo{person}{Yongdong Zhang}, {and} \bibinfo{person}{Meng Wang}.} \bibinfo{year}{2020}\natexlab{}.
\newblock \showarticletitle{Lightgcn: Simplifying and powering graph convolution network for recommendation}. In \bibinfo{booktitle}{\emph{Proceedings of the 43rd International ACM SIGIR conference on research and development in Information Retrieval}}. \bibinfo{pages}{639--648}.
\newblock


\bibitem[He et~al\mbox{.}(2017)]%
        {ncf}
\bibfield{author}{\bibinfo{person}{Xiangnan He}, \bibinfo{person}{Lizi Liao}, \bibinfo{person}{Hanwang Zhang}, \bibinfo{person}{Liqiang Nie}, \bibinfo{person}{Xia Hu}, {and} \bibinfo{person}{Tat-Seng Chua}.} \bibinfo{year}{2017}\natexlab{}.
\newblock \showarticletitle{Neural collaborative filtering}. In \bibinfo{booktitle}{\emph{Proceedings of the 26th international conference on world wide web}}. \bibinfo{pages}{173--182}.
\newblock


\bibitem[Hu et~al\mbox{.}(2024)]%
        {hu2024exact}
\bibfield{author}{\bibinfo{person}{Zhiyu Hu}, \bibinfo{person}{Yang Zhang}, \bibinfo{person}{Minghao Xiao}, \bibinfo{person}{Wenjie Wang}, \bibinfo{person}{Fuli Feng}, {and} \bibinfo{person}{Xiangnan He}.} \bibinfo{year}{2024}\natexlab{}.
\newblock \showarticletitle{Exact and Efficient Unlearning for Large Language Model-based Recommendation}.
\newblock \bibinfo{journal}{\emph{arXiv preprint arXiv:2404.10327}} (\bibinfo{year}{2024}).
\newblock


\bibitem[Huang et~al\mbox{.}(2021)]%
        {huang2021ema}
\bibfield{author}{\bibinfo{person}{Yangsibo Huang}, \bibinfo{person}{Xiaoxiao Li}, {and} \bibinfo{person}{Kai Li}.} \bibinfo{year}{2021}\natexlab{}.
\newblock \showarticletitle{Ema: Auditing data removal from trained models}. In \bibinfo{booktitle}{\emph{Medical Image Computing and Computer Assisted Intervention--MICCAI 2021: 24th International Conference, Strasbourg, France, September 27--October 1, 2021, Proceedings, Part V}}. Springer, \bibinfo{pages}{793--803}.
\newblock


\bibitem[Juneja and Mitra(2021)]%
        {misinformation-chi}
\bibfield{author}{\bibinfo{person}{Prerna Juneja} {and} \bibinfo{person}{Tanushree Mitra}.} \bibinfo{year}{2021}\natexlab{}.
\newblock \showarticletitle{Auditing E-Commerce Platforms for Algorithmically Curated Vaccine Misinformation}. In \bibinfo{booktitle}{\emph{{CHI} '21: {CHI} Conference on Human Factors in Computing Systems}}. \bibinfo{publisher}{{ACM}}, \bibinfo{pages}{186:1--186:27}.
\newblock


\bibitem[Kingma and Ba(2014)]%
        {adam}
\bibfield{author}{\bibinfo{person}{Diederik~P Kingma} {and} \bibinfo{person}{Jimmy Ba}.} \bibinfo{year}{2014}\natexlab{}.
\newblock \showarticletitle{Adam: A method for stochastic optimization}.
\newblock \bibinfo{journal}{\emph{arXiv preprint arXiv:1412.6980}} (\bibinfo{year}{2014}).
\newblock


\bibitem[Koh and Liang(2017)]%
        {IF}
\bibfield{author}{\bibinfo{person}{Pang~Wei Koh} {and} \bibinfo{person}{Percy Liang}.} \bibinfo{year}{2017}\natexlab{}.
\newblock \showarticletitle{Understanding black-box predictions via influence functions}. In \bibinfo{booktitle}{\emph{International conference on machine learning}}. PMLR, \bibinfo{pages}{1885--1894}.
\newblock


\bibitem[Koren et~al\mbox{.}(2009)]%
        {koren2009matrix}
\bibfield{author}{\bibinfo{person}{Yehuda Koren}, \bibinfo{person}{Robert Bell}, {and} \bibinfo{person}{Chris Volinsky}.} \bibinfo{year}{2009}\natexlab{}.
\newblock \showarticletitle{Matrix factorization techniques for recommender systems}.
\newblock \bibinfo{journal}{\emph{Computer}} \bibinfo{volume}{42}, \bibinfo{number}{8} (\bibinfo{year}{2009}), \bibinfo{pages}{30--37}.
\newblock


\bibitem[Koren et~al\mbox{.}(2022)]%
        {CF-survey}
\bibfield{author}{\bibinfo{person}{Yehuda Koren}, \bibinfo{person}{Steffen Rendle}, {and} \bibinfo{person}{Robert Bell}.} \bibinfo{year}{2022}\natexlab{}.
\newblock \showarticletitle{Advances in collaborative filtering}.
\newblock \bibinfo{journal}{\emph{Recommender systems handbook}} (\bibinfo{year}{2022}), \bibinfo{pages}{91--142}.
\newblock


\bibitem[Li et~al\mbox{.}(2024)]%
        {LASER}
\bibfield{author}{\bibinfo{person}{Yuyuan Li}, \bibinfo{person}{Chaochao Chen}, \bibinfo{person}{Xiaolin Zheng}, \bibinfo{person}{Junlin Liu}, {and} \bibinfo{person}{Jun Wang}.} \bibinfo{year}{2024}\natexlab{}.
\newblock \showarticletitle{Making recommender systems forget: Learning and unlearning for erasable recommendation}.
\newblock \bibinfo{journal}{\emph{Knowledge-Based Systems}}  \bibinfo{volume}{283} (\bibinfo{year}{2024}), \bibinfo{pages}{111124}.
\newblock


\bibitem[Li et~al\mbox{.}(2023)]%
        {li2023selective}
\bibfield{author}{\bibinfo{person}{Yuyuan Li}, \bibinfo{person}{Chaochao Chen}, \bibinfo{person}{Xiaolin Zheng}, \bibinfo{person}{Yizhao Zhang}, \bibinfo{person}{Biao Gong}, \bibinfo{person}{Jun Wang}, {and} \bibinfo{person}{Linxun Chen}.} \bibinfo{year}{2023}\natexlab{}.
\newblock \showarticletitle{Selective and collaborative influence function for efficient recommendation unlearning}.
\newblock \bibinfo{journal}{\emph{Expert Systems with Applications}}  \bibinfo{volume}{234} (\bibinfo{year}{2023}), \bibinfo{pages}{121025}.
\newblock


\bibitem[Lin et~al\mbox{.}(2023)]%
        {lin2023can}
\bibfield{author}{\bibinfo{person}{Jianghao Lin}, \bibinfo{person}{Xinyi Dai}, \bibinfo{person}{Yunjia Xi}, \bibinfo{person}{Weiwen Liu}, \bibinfo{person}{Bo Chen}, \bibinfo{person}{Xiangyang Li}, \bibinfo{person}{Chenxu Zhu}, \bibinfo{person}{Huifeng Guo}, \bibinfo{person}{Yong Yu}, \bibinfo{person}{Ruiming Tang}, {et~al\mbox{.}}} \bibinfo{year}{2023}\natexlab{}.
\newblock \showarticletitle{How Can Recommender Systems Benefit from Large Language Models: A Survey}.
\newblock \bibinfo{journal}{\emph{arXiv preprint arXiv:2306.05817}} (\bibinfo{year}{2023}).
\newblock


\bibitem[Liu et~al\mbox{.}(2022)]%
        {liu2022forgetting}
\bibfield{author}{\bibinfo{person}{Wenyan Liu}, \bibinfo{person}{Juncheng Wan}, \bibinfo{person}{Xiaoling Wang}, \bibinfo{person}{Weinan Zhang}, \bibinfo{person}{Dell Zhang}, {and} \bibinfo{person}{Hang Li}.} \bibinfo{year}{2022}\natexlab{}.
\newblock \showarticletitle{Forgetting Fast in Recommender Systems}.
\newblock \bibinfo{journal}{\emph{arXiv preprint arXiv:2208.06875}} (\bibinfo{year}{2022}).
\newblock


\bibitem[Mantelero(2013)]%
        {GDPR}
\bibfield{author}{\bibinfo{person}{Alessandro Mantelero}.} \bibinfo{year}{2013}\natexlab{}.
\newblock \showarticletitle{The EU Proposal for a General Data Protection Regulation and the roots of the ‘right to be forgotten’}.
\newblock \bibinfo{journal}{\emph{Computer Law \& Security Review}} \bibinfo{volume}{29}, \bibinfo{number}{3} (\bibinfo{year}{2013}), \bibinfo{pages}{229--235}.
\newblock


\bibitem[Marchant et~al\mbox{.}(2022)]%
        {marchant2022hard}
\bibfield{author}{\bibinfo{person}{Neil~G Marchant}, \bibinfo{person}{Benjamin~IP Rubinstein}, {and} \bibinfo{person}{Scott Alfeld}.} \bibinfo{year}{2022}\natexlab{}.
\newblock \showarticletitle{Hard to forget: Poisoning attacks on certified machine unlearning}. In \bibinfo{booktitle}{\emph{Proceedings of the AAAI Conference on Artificial Intelligence}}, Vol.~\bibinfo{volume}{36}. \bibinfo{pages}{7691--7700}.
\newblock


\bibitem[Martinez et~al\mbox{.}(2023)]%
        {martinez2023approximating}
\bibfield{author}{\bibinfo{person}{Javier~Abad Martinez}, \bibinfo{person}{Umang Bhatt}, \bibinfo{person}{Adrian Weller}, {and} \bibinfo{person}{Giovanni Cherubin}.} \bibinfo{year}{2023}\natexlab{}.
\newblock \showarticletitle{Approximating Full Conformal Prediction at Scale via Influence Functions}. In \bibinfo{booktitle}{\emph{Proceedings of the AAAI Conference on Artificial Intelligence}}, Vol.~\bibinfo{volume}{37}. \bibinfo{pages}{6631--6639}.
\newblock


\bibitem[Milano et~al\mbox{.}(2020)]%
        {rec-ethical}
\bibfield{author}{\bibinfo{person}{Silvia Milano}, \bibinfo{person}{Mariarosaria Taddeo}, {and} \bibinfo{person}{Luciano Floridi}.} \bibinfo{year}{2020}\natexlab{}.
\newblock \showarticletitle{Recommender systems and their ethical challenges}.
\newblock \bibinfo{journal}{\emph{{AI} Soc.}} \bibinfo{volume}{35}, \bibinfo{number}{4} (\bibinfo{year}{2020}), \bibinfo{pages}{957--967}.
\newblock
\urldef\tempurl%
\url{https://doi.org/10.1007/s00146-020-00950-y}
\showDOI{\tempurl}


\bibitem[Nguyen et~al\mbox{.}(2022)]%
        {nguyen2022survey}
\bibfield{author}{\bibinfo{person}{Thanh~Tam Nguyen}, \bibinfo{person}{Thanh~Trung Huynh}, \bibinfo{person}{Phi~Le Nguyen}, \bibinfo{person}{Alan Wee-Chung Liew}, \bibinfo{person}{Hongzhi Yin}, {and} \bibinfo{person}{Quoc Viet~Hung Nguyen}.} \bibinfo{year}{2022}\natexlab{}.
\newblock \showarticletitle{A Survey of Machine Unlearning}.
\newblock \bibinfo{journal}{\emph{arXiv preprint arXiv:2209.02299}} (\bibinfo{year}{2022}).
\newblock


\bibitem[Nguyen{-}Duc et~al\mbox{.}(2023)]%
        {Nguyen-DucTTHND23}
\bibfield{author}{\bibinfo{person}{Thang Nguyen{-}Duc}, \bibinfo{person}{Hoang Thanh{-}Tung}, \bibinfo{person}{Quan~Hung Tran}, \bibinfo{person}{Dang Huu{-}Tien}, \bibinfo{person}{Hieu Nguyen}, \bibinfo{person}{Anh T.~V. Dau}, {and} \bibinfo{person}{Nghi Bui}.} \bibinfo{year}{2023}\natexlab{}.
\newblock \showarticletitle{Class based Influence Functions for Error Detection}. In \bibinfo{booktitle}{\emph{Proceedings of the 61st Annual Meeting of the Association for Computational Linguistics (Volume 2: Short Papers), {ACL} 2023}}. \bibinfo{pages}{1204--1218}.
\newblock


\bibitem[Onokoy and Lavendels(2019)]%
        {modern-recsys}
\bibfield{author}{\bibinfo{person}{Lyudmila Onokoy} {and} \bibinfo{person}{Jurijs Lavendels}.} \bibinfo{year}{2019}\natexlab{}.
\newblock \showarticletitle{Modern Approaches to Building Recommender Systems for Online Stores}.
\newblock \bibinfo{journal}{\emph{Appl. Comput. Syst.}} \bibinfo{volume}{24}, \bibinfo{number}{1} (\bibinfo{year}{2019}), \bibinfo{pages}{18--24}.
\newblock


\bibitem[Rendle(2010)]%
        {rendleFM}
\bibfield{author}{\bibinfo{person}{Steffen Rendle}.} \bibinfo{year}{2010}\natexlab{}.
\newblock \showarticletitle{Factorization machines}. In \bibinfo{booktitle}{\emph{2010 IEEE International conference on data mining}}. IEEE, \bibinfo{pages}{995--1000}.
\newblock


\bibitem[Rendle et~al\mbox{.}(2009)]%
        {rendle2012bpr}
\bibfield{author}{\bibinfo{person}{Steffen Rendle}, \bibinfo{person}{Christoph Freudenthaler}, \bibinfo{person}{Zeno Gantner}, {and} \bibinfo{person}{Lars Schmidt-Thieme}.} \bibinfo{year}{2009}\natexlab{}.
\newblock \showarticletitle{BPR: Bayesian personalized ranking from implicit feedback}. In \bibinfo{booktitle}{\emph{Proceedings of the Twenty-Fifth Conference on Uncertainty in Artificial Intelligence}}. \bibinfo{pages}{452–461}.
\newblock


\bibitem[Schelter(2020)]%
        {schelter8364amnesia}
\bibfield{author}{\bibinfo{person}{Sebastian Schelter}.} \bibinfo{year}{2020}\natexlab{}.
\newblock \showarticletitle{Amnesia-a selection of machine learning models that can forget user data very fast}.
\newblock \bibinfo{journal}{\emph{suicide}} \bibinfo{volume}{8364}, \bibinfo{number}{44035} (\bibinfo{year}{2020}), \bibinfo{pages}{46992}.
\newblock


\bibitem[Shan et~al\mbox{.}(2020)]%
        {shan2020fawkes}
\bibfield{author}{\bibinfo{person}{Shawn Shan}, \bibinfo{person}{Emily Wenger}, \bibinfo{person}{Jiayun Zhang}, \bibinfo{person}{Huiying Li}, \bibinfo{person}{Haitao Zheng}, {and} \bibinfo{person}{Ben~Y Zhao}.} \bibinfo{year}{2020}\natexlab{}.
\newblock \showarticletitle{Fawkes: Protecting privacy against unauthorized deep learning models}. In \bibinfo{booktitle}{\emph{29th USENIX security symposium (USENIX Security 20)}}. \bibinfo{pages}{1589--1604}.
\newblock


\bibitem[Tak{\'a}cs et~al\mbox{.}(2011)]%
        {takacs2011applications}
\bibfield{author}{\bibinfo{person}{G{\'a}bor Tak{\'a}cs}, \bibinfo{person}{Istv{\'a}n Pil{\'a}szy}, {and} \bibinfo{person}{Domonkos Tikk}.} \bibinfo{year}{2011}\natexlab{}.
\newblock \showarticletitle{Applications of the conjugate gradient method for implicit feedback collaborative filtering}. In \bibinfo{booktitle}{\emph{Proceedings of the fifth ACM conference on Recommender systems}}. \bibinfo{pages}{297--300}.
\newblock


\bibitem[Tarun et~al\mbox{.}(2023)]%
        {tarun2021fast}
\bibfield{author}{\bibinfo{person}{Ayush~K Tarun}, \bibinfo{person}{Vikram~S Chundawat}, \bibinfo{person}{Murari Mandal}, {and} \bibinfo{person}{Mohan Kankanhalli}.} \bibinfo{year}{2023}\natexlab{}.
\newblock \showarticletitle{Fast yet effective machine unlearning}.
\newblock \bibinfo{journal}{\emph{IEEE Transactions on Neural Networks and Learning Systems}} (\bibinfo{year}{2023}).
\newblock


\bibitem[Thudi et~al\mbox{.}(2022a)]%
        {thudi2022unrolling}
\bibfield{author}{\bibinfo{person}{Anvith Thudi}, \bibinfo{person}{Gabriel Deza}, \bibinfo{person}{Varun Chandrasekaran}, {and} \bibinfo{person}{Nicolas Papernot}.} \bibinfo{year}{2022}\natexlab{a}.
\newblock \showarticletitle{Unrolling sgd: Understanding factors influencing machine unlearning}. In \bibinfo{booktitle}{\emph{2022 IEEE 7th European Symposium on Security and Privacy (EuroS\&P)}}. IEEE, \bibinfo{pages}{303--319}.
\newblock


\bibitem[Thudi et~al\mbox{.}(2022b)]%
        {thudi2022necessity}
\bibfield{author}{\bibinfo{person}{Anvith Thudi}, \bibinfo{person}{Hengrui Jia}, \bibinfo{person}{Ilia Shumailov}, {and} \bibinfo{person}{Nicolas Papernot}.} \bibinfo{year}{2022}\natexlab{b}.
\newblock \showarticletitle{On the necessity of auditable algorithmic definitions for machine unlearning}. In \bibinfo{booktitle}{\emph{31st USENIX Security Symposium (USENIX Security 22)}}. \bibinfo{pages}{4007--4022}.
\newblock


\bibitem[Tommasel and Menczer(2022)]%
        {misinformation-recsys}
\bibfield{author}{\bibinfo{person}{Antonela Tommasel} {and} \bibinfo{person}{Filippo Menczer}.} \bibinfo{year}{2022}\natexlab{}.
\newblock \showarticletitle{Do Recommender Systems Make Social Media More Susceptible to Misinformation Spreaders?}. In \bibinfo{booktitle}{\emph{Proceedings of the 16th ACM Conference on Recommender Systems}}. \bibinfo{publisher}{Association for Computing Machinery}, \bibinfo{pages}{550–555}.
\newblock
\showISBNx{9781450392785}


\bibitem[Warnecke et~al\mbox{.}(2023)]%
        {inf-unlearn}
\bibfield{author}{\bibinfo{person}{Alexander Warnecke}, \bibinfo{person}{Lukas Pirch}, \bibinfo{person}{Christian Wressnegger}, {and} \bibinfo{person}{Konrad Rieck}.} \bibinfo{year}{2023}\natexlab{}.
\newblock \showarticletitle{Machine Unlearning of Features and Labels}. In \bibinfo{booktitle}{\emph{Network and Distributed System Security Symposium (NDSS) 2023}}.
\newblock


\bibitem[Wu et~al\mbox{.}(2021)]%
        {wu2021triple}
\bibfield{author}{\bibinfo{person}{Chenwang Wu}, \bibinfo{person}{Defu Lian}, \bibinfo{person}{Yong Ge}, \bibinfo{person}{Zhihao Zhu}, {and} \bibinfo{person}{Enhong Chen}.} \bibinfo{year}{2021}\natexlab{}.
\newblock \showarticletitle{Triple Adversarial Learning for Influence based Poisoning Attack in Recommender Systems}. In \bibinfo{booktitle}{\emph{Proceedings of the 27th ACM SIGKDD Conference on Knowledge Discovery \& Data Mining}}. \bibinfo{pages}{1830--1840}.
\newblock


\bibitem[Wu et~al\mbox{.}(2022)]%
        {PUMA}
\bibfield{author}{\bibinfo{person}{Ga Wu}, \bibinfo{person}{Masoud Hashemi}, {and} \bibinfo{person}{Christopher Srinivasa}.} \bibinfo{year}{2022}\natexlab{}.
\newblock \showarticletitle{{PUMA:} Performance Unchanged Model Augmentation for Training Data Removal}. In \bibinfo{booktitle}{\emph{Thirty-Sixth {AAAI} Conference on Artificial Intelligence}}. \bibinfo{publisher}{{AAAI} Press}, \bibinfo{pages}{8675--8682}.
\newblock


\bibitem[Wu et~al\mbox{.}(2023a)]%
        {GIF}
\bibfield{author}{\bibinfo{person}{Jiancan Wu}, \bibinfo{person}{Yi Yang}, \bibinfo{person}{Yuchun Qian}, \bibinfo{person}{Yongduo Sui}, \bibinfo{person}{Xiang Wang}, {and} \bibinfo{person}{Xiangnan He}.} \bibinfo{year}{2023}\natexlab{a}.
\newblock \showarticletitle{GIF: A General Graph Unlearning Strategy via Influence Function}. In \bibinfo{booktitle}{\emph{Proceedings of the ACM Web Conference 2023}}. \bibinfo{pages}{651–661}.
\newblock


\bibitem[Wu et~al\mbox{.}(2023b)]%
        {wu2023survey}
\bibfield{author}{\bibinfo{person}{Likang Wu}, \bibinfo{person}{Zhi Zheng}, \bibinfo{person}{Zhaopeng Qiu}, \bibinfo{person}{Hao Wang}, \bibinfo{person}{Hongchao Gu}, \bibinfo{person}{Tingjia Shen}, \bibinfo{person}{Chuan Qin}, \bibinfo{person}{Chen Zhu}, \bibinfo{person}{Hengshu Zhu}, \bibinfo{person}{Qi Liu}, {et~al\mbox{.}}} \bibinfo{year}{2023}\natexlab{b}.
\newblock \showarticletitle{A Survey on Large Language Models for Recommendation}.
\newblock \bibinfo{journal}{\emph{arXiv preprint arXiv:2305.19860}} (\bibinfo{year}{2023}).
\newblock


\bibitem[Xu et~al\mbox{.}(2023b)]%
        {survey2}
\bibfield{author}{\bibinfo{person}{Heng Xu}, \bibinfo{person}{Tianqing Zhu*}, \bibinfo{person}{Lefeng Zhang}, \bibinfo{person}{Wanlei Zhou}, {and} \bibinfo{person}{Philip~S. Yu}.} \bibinfo{year}{2023}\natexlab{b}.
\newblock \showarticletitle{Machine Unlearning: A Survey}.
\newblock \bibinfo{journal}{\emph{ACM Comput. Surv.}} (\bibinfo{year}{2023}).
\newblock
\newblock
\shownote{Just Accepted}.


\bibitem[Xu et~al\mbox{.}(2023a)]%
        {xu2023netflix}
\bibfield{author}{\bibinfo{person}{Mimee Xu}, \bibinfo{person}{Jiankai Sun}, \bibinfo{person}{Xin Yang}, \bibinfo{person}{Kevin Yao}, {and} \bibinfo{person}{Chong Wang}.} \bibinfo{year}{2023}\natexlab{a}.
\newblock \showarticletitle{Netflix and Forget: Efficient and Exact Machine Unlearning from Bi-linear Recommendations}.
\newblock \bibinfo{journal}{\emph{arXiv preprint arXiv:2302.06676}} (\bibinfo{year}{2023}).
\newblock


\bibitem[Yi et~al\mbox{.}(2014)]%
        {yi2014robust}
\bibfield{author}{\bibinfo{person}{Huawei Yi}, \bibinfo{person}{Fuzhi Zhang}, {and} \bibinfo{person}{Jie Lan}.} \bibinfo{year}{2014}\natexlab{}.
\newblock \showarticletitle{A robust collaborative recommendation algorithm based on k-distance and Tukey M-estimator}.
\newblock \bibinfo{journal}{\emph{China Communications}} \bibinfo{volume}{11}, \bibinfo{number}{9} (\bibinfo{year}{2014}), \bibinfo{pages}{112--123}.
\newblock


\bibitem[Yu et~al\mbox{.}(2020)]%
        {yu2020influence}
\bibfield{author}{\bibinfo{person}{Jiangxing Yu}, \bibinfo{person}{Hong Zhu}, \bibinfo{person}{Chih-Yao Chang}, \bibinfo{person}{Xinhua Feng}, \bibinfo{person}{Bowen Yuan}, \bibinfo{person}{Xiuqiang He}, {and} \bibinfo{person}{Zhenhua Dong}.} \bibinfo{year}{2020}\natexlab{}.
\newblock \showarticletitle{Influence function for unbiased recommendation}. In \bibinfo{booktitle}{\emph{Proceedings of the 43rd International ACM SIGIR Conference on Research and Development in Information Retrieval}}. \bibinfo{pages}{1929--1932}.
\newblock


\bibitem[Yuan et~al\mbox{.}(2022)]%
        {yuan2022federated}
\bibfield{author}{\bibinfo{person}{Wei Yuan}, \bibinfo{person}{Hongzhi Yin}, \bibinfo{person}{Fangzhao Wu}, \bibinfo{person}{Shijie Zhang}, \bibinfo{person}{Tieke He}, {and} \bibinfo{person}{Hao Wang}.} \bibinfo{year}{2022}\natexlab{}.
\newblock \showarticletitle{Federated Unlearning for On-Device Recommendation}.
\newblock \bibinfo{journal}{\emph{arXiv preprint arXiv:2210.10958}} (\bibinfo{year}{2022}).
\newblock


\bibitem[Zeng et~al\mbox{.}(2022)]%
        {unlean-app}
\bibfield{author}{\bibinfo{person}{Yi Zeng}, \bibinfo{person}{Si Chen}, \bibinfo{person}{Won Park}, \bibinfo{person}{Zhuoqing Mao}, \bibinfo{person}{Ming Jin}, {and} \bibinfo{person}{Ruoxi Jia}.} \bibinfo{year}{2022}\natexlab{}.
\newblock \showarticletitle{Adversarial Unlearning of Backdoors via Implicit Hypergradient}. In \bibinfo{booktitle}{\emph{The Tenth International Conference on Learning Representations}}. \bibinfo{publisher}{OpenReview.net}.
\newblock


\bibitem[Zhang et~al\mbox{.}(2020b)]%
        {zhang2020practical}
\bibfield{author}{\bibinfo{person}{Hengtong Zhang}, \bibinfo{person}{Yaliang Li}, \bibinfo{person}{Bolin Ding}, {and} \bibinfo{person}{Jing Gao}.} \bibinfo{year}{2020}\natexlab{b}.
\newblock \showarticletitle{Practical data poisoning attack against next-item recommendation}. In \bibinfo{booktitle}{\emph{Proceedings of The Web Conference 2020}}. \bibinfo{pages}{2458--2464}.
\newblock


\bibitem[Zhang et~al\mbox{.}(2019)]%
        {deep-recsys}
\bibfield{author}{\bibinfo{person}{Shuai Zhang}, \bibinfo{person}{Lina Yao}, \bibinfo{person}{Aixin Sun}, {and} \bibinfo{person}{Yi Tay}.} \bibinfo{year}{2019}\natexlab{}.
\newblock \showarticletitle{Deep Learning Based Recommender System: {A} Survey and New Perspectives}.
\newblock \bibinfo{journal}{\emph{{ACM} Comput. Surv.}} \bibinfo{volume}{52}, \bibinfo{number}{1} (\bibinfo{year}{2019}), \bibinfo{pages}{5:1--5:38}.
\newblock


\bibitem[Zhang et~al\mbox{.}(2021)]%
        {zhang2021sample}
\bibfield{author}{\bibinfo{person}{Wei Zhang}, \bibinfo{person}{Ziming Huang}, \bibinfo{person}{Yada Zhu}, \bibinfo{person}{Guangnan Ye}, \bibinfo{person}{Xiaodong Cui}, {and} \bibinfo{person}{Fan Zhang}.} \bibinfo{year}{2021}\natexlab{}.
\newblock \showarticletitle{On sample based explanation methods for NLP: Faithfulness, efficiency and semantic evaluation}. In \bibinfo{booktitle}{\emph{Proceedings of the 59th Annual Meeting of the Association for Computational Linguistics and the 11th International Joint Conference on Natural Language Processing (Volume 1: Long Papers)}}. \bibinfo{pages}{5399--5411}.
\newblock


\bibitem[Zhang et~al\mbox{.}(2020a)]%
        {SML}
\bibfield{author}{\bibinfo{person}{Yang Zhang}, \bibinfo{person}{Fuli Feng}, \bibinfo{person}{Chenxu Wang}, \bibinfo{person}{Xiangnan He}, \bibinfo{person}{Meng Wang}, \bibinfo{person}{Yan Li}, {and} \bibinfo{person}{Yongdong Zhang}.} \bibinfo{year}{2020}\natexlab{a}.
\newblock \showarticletitle{How to retrain recommender system? A sequential meta-learning method}. In \bibinfo{booktitle}{\emph{Proceedings of the 43rd International ACM SIGIR Conference on Research and Development in Information Retrieval}}. \bibinfo{pages}{1479--1488}.
\newblock


\bibitem[Zhang et~al\mbox{.}(2023)]%
        {DIL}
\bibfield{author}{\bibinfo{person}{Yang Zhang}, \bibinfo{person}{Tianhao Shi}, \bibinfo{person}{Fuli Feng}, \bibinfo{person}{Wenjie Wang}, \bibinfo{person}{Dingxian Wang}, \bibinfo{person}{Xiangnan He}, {and} \bibinfo{person}{Yongdong Zhang}.} \bibinfo{year}{2023}\natexlab{}.
\newblock \showarticletitle{Reformulating CTR Prediction: Learning Invariant Feature Interactions for Recommendation}. In \bibinfo{booktitle}{\emph{Proceedings of the 46th International ACM SIGIR Conference on Research and Development in Information Retrieval}}.
\newblock


\end{thebibliography}

\appendix

\end{document}